\documentclass[a4paper]{article}

\usepackage{cite}      

\usepackage{graphicx}  

\usepackage{subfigure} 

\usepackage{amsmath}   

\newcommand{\1}{|}
\newcommand{\0}{\bigcirc}

\newcommand{\f}{\bot}
\newcommand{\+}{\land}
\newcommand{\Or}{\lor}
\newcommand{\Not}{\neg}
\newcommand{\goesto}{\longrightarrow}
\newcommand{\IMP}{\to}

\newcommand{\NXOR}{\leftrightarrow}
\newcommand{\NAND}{\bar{\land}}
\newcommand{\NOR}{\bar{\lor}}
\newcommand{\XOR}{\not \leftrightarrow}

\begin{document}

\title{Single Memristor Logic Gates: From NOT to a Full Adder}

\author{Ella Gale\\
International Center for Unconventional Computing,\\ Bristol Robotics Laboratory\\
Bristol, UK BS16 1QY\\
Current address: Department of Chemistry, \\
University of Bath, Claverton Down,
Bath, UK, BA2 7AY\\
Email: E.Gale@bath.ac.uk}


%


\maketitle

\begin{abstract}
Memristors have been suggested as a novel route to neuromorphic computing based on the similarity between them and neurons (specifically synapses and ion pumps). The d.c. action of the memristor is a current spike which imparts a short-term memory to the device. Here it is demonstrated that this short-term memory works exactly like habituation (e.g. in \emph{Aplysia}). We elucidate the physical rules, based on energy conservation, governing the interaction of these current spikes: summation, `bounce-back', directionality and `diminishing returns'. Using these rules, we introduce 4 different logical systems to implement sequential logic in the memristor and demonstrate how sequential logic works by instantiating a NOT gate, an AND gate, an XOR gate and a Full Adder with a single memristor. The Full Adder makes use of the memristor's short-term memory to add together three binary values and outputs the sum, the carry digit and even the order they were input in. A memristor full adder also outputs the arithmetical sum of bits, allowing for a logically (but not physically) reversible system. Essentially, we can replace an input/output port with an extra time-step, allowing a single memristor to do a hither-to unexpectedly large amount of computation. This makes up for the memristor's slow operation speed and may relate to how neurons do a similarly-large computation with such slow operations speeds. We propose that using spiking logic, either in gates or as neuron-analogues, with plastic rewritable connections between them, would allow the building of a neuromorphic computer.
\end{abstract}


%

\section{Introduction}

What is a memristor? It is a resistor with memory. It is neuron-like. And it is more than theoretical, it is physical device acting in the physical world, which means that it is non-conservative with respects to energy. I shall introduce the subject and explain the contents of this paper working from these three simple statements. 

\paragraph{A resistor with memory}

Chua introduced the memristor concept in 1971~\cite{14} as a device which related its internal state to the time-integral of current or voltage. The concept has been expanded and refined~\cite{72,84,119} and there is still some debate about what definition should be used to describe the memristor (see this recent review~\cite{RevMemReRAM}) although little of that concerns us here. 

In the memristor, a.c. measurements give rise to a distinctive pinched hysteresis loop (taken to be one of the fingerprints of the device~\cite{222}), where the memory is encoded in the hysteresis: reading off a current at a specific voltage we see two different currents because one has been taken up to the maximum voltage and back down again and the other has not. The memristor memory is usually stored in ions, for example in the best-known version of the device~\cite{15}, it is stored in oxygen vacancies, this is also the case in our devices\cite{M0,260}. Putting a voltage across the device causes the movement of ions which changes the resistance, leading to an altered current (there is far more detail available than this, see\cite{RevMemReRAM,296,155}). 

My work has involved looking at the `d.c.' properties of the memristor: how it reacts to a steady non-varying voltage or a series of pulses. It was found that when a voltage is applied in this way there is a resultant current spike which has been seen is our~\cite{hystc,SpcJ} and other's~\cite{123,macro1,261} devices.

Memory allows the association of events which happen (and have happened) at different points in time. In this paper, I will present how this allows the device to store data temporarily and allow input bits to interact through time see sections~\ref{sec:rules,sec:examples}. This idea is essentially the opposite of parallelisation, when building parallel circuitry we are essentially replacing the time (as in time for a computation) with space (extra processors). Here we are replacing space (components on a circuit board) with time (increased number of steps) using the memory to store the computation through time. The gates made using this motivates the question of what is the circuit and logical complexity of such a mode of operation; essentially what is our conversion rate between space and time? We need to know the trade-offs to allow us to get the maximum efficiency for a circuit. For example, we are reaching the limits of how far we can continue shrinking computers in accordance with Moore's law~\cite{MooresLaw}. To overcome this, the International Technology Roadmap for Semiconductors suggested the approach of `More-than-Moore' where the components are capable of performing more functions~\cite{322}. To plan a circuit with these components we would need to precisely compare and contrast this extra functionality.

\paragraph{Neuron-like}

Memristors have been compared to both neurons~\cite{84,247,248} and synapses~\cite{41,217}, and as such, have been enthusiastically received by the neuromorphic computing community. The first experimental memristor paper~\cite{15} suggested that because memristors combined processing and memory in the same component, they were similar to neurons and could be the basis of a brain-like computer. This same group later presented the `neuristor'~\cite{272} -- a combination of memristors and capacitors that presented repetitive brain-like dynamics. 

Brain dynamics are known to be chaotic and it has been suggested that neurons are poised at the edge of chaos~\cite{248}, so the search for chaos using memristors (a search also inspired by the similarity of the memristor's operation to the Chua diode in Chua's circuit~\cite{256,257} -- the simplest chaotic circuit) is relevant to us here. From simulations, it has been shown that a Chua circuit can be built including a memristor~\cite{252,82,70,61,232}. Experimentally, chaos-like dynamics have been observed in memristor circuits~\cite{EllaMattia} and chaos has been demonstrated as arising from a single memristor~\cite{EllaC1}.

One of Chua's early papers~\cite{84} suggested that memristors were present in the Hodgkin Huxely~\cite{HH} model~\footnote{The Hodgkin-Huxley model describes an electrical circuit equivalent for a neuron membrane. A neuron, like all cells, consists of a cell membrane that separates the inside of the cell from the outside. The cell membrane contains proteins called `protein pumps' that pump ions out of the neuron doing work against a chemical gradient. When the neuron fires, the ions flood back in, giving a quick change in charge that leads the voltage spike that transmits the electrical signal. In the Hodgkin-Huxley model the cell membrane is well described as a capacitor and the protein pumps were described as time-varying resistors and the power provided by respiration was represented as a battery. All the components of the model could be bought in a hardware store, but the time-varying resistors, used to model protein pumps, did not really have a hardware analogue.}, a statement he later developed~\cite{ChuaNanotech,247,248}. The ion pumps in neural cell membranes were modelled as time-varying resistors, the existence of which suggested that biological brains should have huge impedances (and resulting power draw) which we know they do not have. By replacing the time-varying resistor with time-invarient memristors running off the time-varying ion concentration, the model no longer requires unnatural 
impedences~\cite{ChuaNanotech}. 

Of interest to us is another property of neurons: habituation. Habituation~\cite{SquireMemoryBook} is the the learned ability to ignore unnecessary stmulii, such as an author not attending to the surrounding background noises of a coffee shop when writing (incidentally, as this learning is unconscious, it takes a great deal of mental training to turn it off, although long-time meditators have been shown to be able to do this~\cite{374}). One of the first and most famous habituation experiments involved the `sea snail' (confusingly also known as `sea hare') or \emph{aplysia}. This invertebrate has a sensitive feeding tube which it doesn't want to get damaged and so will withdraw if the tube is stimulated (like a child will stop sticking its tongue out if you threaten to grab it). In an experiment the feeding tube was lightly brushed with a delicate paintbrush (a harmless intervention), and, over time, the \emph{aplysia} learned not to attend to this stimulus and not to withdraw its tongue. \emph{Aplysia} is a simple creature with only 20,000 neurons, and the neural pathways for this response has been mapped out (and requires only 4 neurons). The response at the sensory neuron does not diminish, but with training the corresponding motor response signals (recorded at motor neuron $L7_{g}$) decrease in size and this is the physiological cause of habituation. 

Chua demonstrated theoretically that an ideal memristor stimulated by repeated voltage spikes would alter its memristive state and present a decreasing current response spike (see figure 7 in~\cite{ChuaNanotech}) and this result strengthened previous theoretical work demonstrating that the neural protein pumps in the Hodgkin-Huxley model fo the neuron~\cite{HH} was best described theoretically as a memristor~\cite{247,248}. We shall replicate this theoretical result with a real memristor in section~\ref{sec:Bio}.

The advantages of using spike interactions are many-fold. The memristor switching itself can be slow~\cite{260} but the spikes can interact much faster, the output of which is `held' in the short-term memory of the memristor, which gives rise to, if not faster processing, more complex operations within a given time-frame than is usually the case in standard electronics.

\paragraph{A real device}

The first experimental memristor paper~\cite{15} was published in 2008 and concerned a titanium dioxide memristor and made the link to Chua's theoretical device. However the field of ReRAM which pre-dated this discovery by around 20 years offered many other chemistries that did the same thing and expanded our understanding of the memrsitor's operation via their materials science models (the interested reader is referred to the following reviews:~\cite{RevMemReRAM,155}).  

Unconventional computing~\cite{adamatzky2007unconventional,adamatzky2006utopian} (also sometimes called natural computation~\cite{gheorghe2005molecular}) investigates the computation of non-traditional systems, such as biological computers (like Slime mould~\cite{AndysBook}), bio-inspired computers (like C.A.)~\cite{adamatzky2001computing}, chemical computing (like the B-Z reaction)~\cite{adamatzky2001computing}, resevoir computing, quantum computing, billiard-ball computing (conservative computation). 

Knowledge of unconventional computing inspires us to consider the computer as a physical system which is instantiated in the real world. If we look at memristors this way, we start to consider the way it uses energy and obeys the laws of physics, such as the conservation of energy. And thus when designing logic gates for these systems we need to consider the physical reality of the device, rather than just an abstraction. This will be covered in section~\ref{sec:rules}.

\subsection{Logic}

To design a circuit, we need to consider circuit complexity. In this paper, we use the Arabic numerals \{1,2,3\} as a set of counting numbers capable of standard arithmetical operations and we use the following set of symbols $\{ \0, \1 \}$ for logical values 0 and 1 in binary.

\subsection{Circuit Complexity}

To decide how to design a circuit board, we need to be able to compare logically equivalent methods of performing an arbitrary logical function. There have been a few approaches to determining how `complex' a circuit is. Circuit size complexity of a Boolean function is the minimal size of a circuit that can compute that function (in the abstract mathematical sense, how many logic gates required, in the concrete electronic engineering sense we would take into account the size of those components). Circuit depth complexity of a Boolean function is the minimal depth of a circuit, i.e. the maximum length of a path from input to output gate for the physical devices. Another approach is the graph of nodes where each node is a logic gate. Karnaugh maps~\cite{377,378} are often used as a method of finding the best approach to build a logical system or express a logical function.



\subsection{Syncronous and asyncronous logic}

In this section the word `logic' refers not to the operations of $\1$s and $\0$s but the methods of combining them. Currently, most logical operations are done in a synchronous manner, the bits are operated on at the same time and a clock pulse is required to keep the information in time (a delay will lead to incorrect values). Asynchronous logic is older in this type of logic the clock pulse in unnecessary.

We also need to briefly cover reversible and irreversible computation. Most computation is irreversible, for example all 2-bit logic gates lose a bit of information and it is this fact that suggested to Shannon that information could be accounted for as an energy and introduced the concept of entropy to computing~\cite{Shannon}. For example if we do a 2-bit OR operation we have no way of telling if a $\1$ output is the result of a $\{ \0, \1\}$, $\{ \1, \0 \}$ or $\{ \1, \1\}$ input. If we can do this, the computation is said to be reversible. Physically reversible computation is extremely rare (examples), logically reversible merelt requires that we can reconstruct the inputs from the outputs.

\subsection{How logic gates have been instantiated with memristors}

This is not the first paper on how to make logic gates with memristors. Strukov et al~\cite{242} used implication logic to design logic gates which required two memrsitors (IMP-FALSE ($\{ \IMP, \f\}$) logic is Turing complete, but somewhat unfamiliar to computer scientists). The most notable Boolean logic gates were simulated by Pershin and di Ventra~\cite{PAndV} and required a memcapacitor, three or four memristive systems and a resistor. Before the gate was sent the two bits of data, a set of initialization pulses were required to be sent to put the gate into the correct state to give the correct answer. This system, however, is not true Boolean logic because these initialization pulses were different dependent on what the logic to follow would be. Thus the gate can not be considered to be operating only on the two bits of input data and is not a simple Boolean logic gate (it is a Turing machine doing a computation on several bits of data (Boolean input pulses and initialization pulses) which is capable of modelling a Boolean logic gate). Note also that this scheme was tested with memristor emulators, not real devices. There have been other more complex designs for memristor based Boolean logic gates, the simplest of which requires 11 circuit elements~\cite{Pino} (and one of the authors, Pino, has several patented memristor logic gates as well). In this paper, we will demonstrate how to perform Boolean logic with a single memristor.

Interesting work involving designing memristor logic gates by Lehtonen and Laiho. In ~\cite{370} they work with implication logic using a memristor reset as the false operation, they also suggest that 3 memrsitors are sufficient to compute any 2-bit in, 1-bit out Boolean function and state that: $2^2 \rightarrow 2^m$ need $m+2$ working memristors, for example a full adder would require 10 memristors under this scheme as it is $2^{2^{3}}$. In ~\cite{371} they suggest using parallelism to avoid the sequential nature of memristors, and choose NcIMP as the best version of Implication logic for memrsitor cross-bars.












This paper starts with the habituation results in section~\ref{sec:Bio} as an example of memristors natively mimicking neural circuits. I then describe the short-term memory of the memristor in section~\ref{sec:repro}, then describe possible methods of performing logic using it in section~\ref{sec:logicmethods}. From there, we are in a position to elucidate the rules for spiking memristor logic~\ref{sec:rules}, and demonstrate them with some examples (section~\ref{sec:examples}), simple logic gates like OR (section~\ref{sec:simplegates}), complex logic gates like XOR (section~\ref{sec:XOR}), before finally describing how to make a logically-reversible full adder in section~\ref{sec:fulladder}. 

\section{Experimental Methodology}

Memristors were fabricated as in~\cite{260} using the TiO$_2$ sol-gel as described in~\cite{M0} and were measured using a Keithley electrometer, with a set time-step of 0.1s, which gives an actual output of 0.16s (time-steps are padded by a settling time to ensure accuracy). After each logical test, the memristor was left for 40 time-steps ($\sim40s$) to lose its short-term memory, i.e. reset to the null state. All presented results plots are experimental data.

Electronic device modelling of a biological habituation learning experiment was performed as follows: 40 0.1V spikes of 0.084s duration were applied 0.084 seconds apart (i.e. it was continuous for 3.35s) which is longer than the duration of the short-term memory of the memristor~\cite{243} to give a baseline non-learning response. Then the device was zeroed for the same amount of time. Then 6 spikes of +0.1V of 0.084s (1 timestep) were applied 1.168 (2 times steps) seconds apart and the current response of the device were recorded at each step (to give the response to the spike on the first step, and the `bounceback' response on the second when the device was zeroed). This approach was repeated with a negative voltage input of -0.01V.

\section{Results}

\subsection{Properties of the memristor\label{sec:properties}}

We shall explain how the memristor works by demonstrating habituation, before elucidating memristor design rules and using them to design circuits.
 
\subsubsection{Recreation of biological experiment using a memristor~\label{sec:Bio}}

A famous study of an amensiac patient H.M. who, as a result of ill-advised brain surgery, had his medial temporal lobe removed to treat his epilepsy. As a result he was not able to convert short term memories into long-term memories (a mental deficiency made famous by the film Memento), however, he was able to learn and improve at certain tasks without ever being aware that he had learnt it, this is the type of `automatic' skills learnt through practice, like martial arts or driving a car~\cite{Mem1,Mem2}. This type of memory is called non-declarative memory and includes `how-to' types of memory, and elementary reflexive (as in it becomes a reflex) learning such as habituation, sensitisation (ability to attend to dangerous stimuli) and classical (association of two stimuli) and operant (association of a subject's actions and a stimuli) conditioning (trained ingrained responses), all of which is stored in the unconscious mind.

\begin{figure}%
\centering
\subfigure[][]{%
\label{fig:Positive}%
 \includegraphics[bb=0 0 576 432,scale=0.3,keepaspectratio=true]{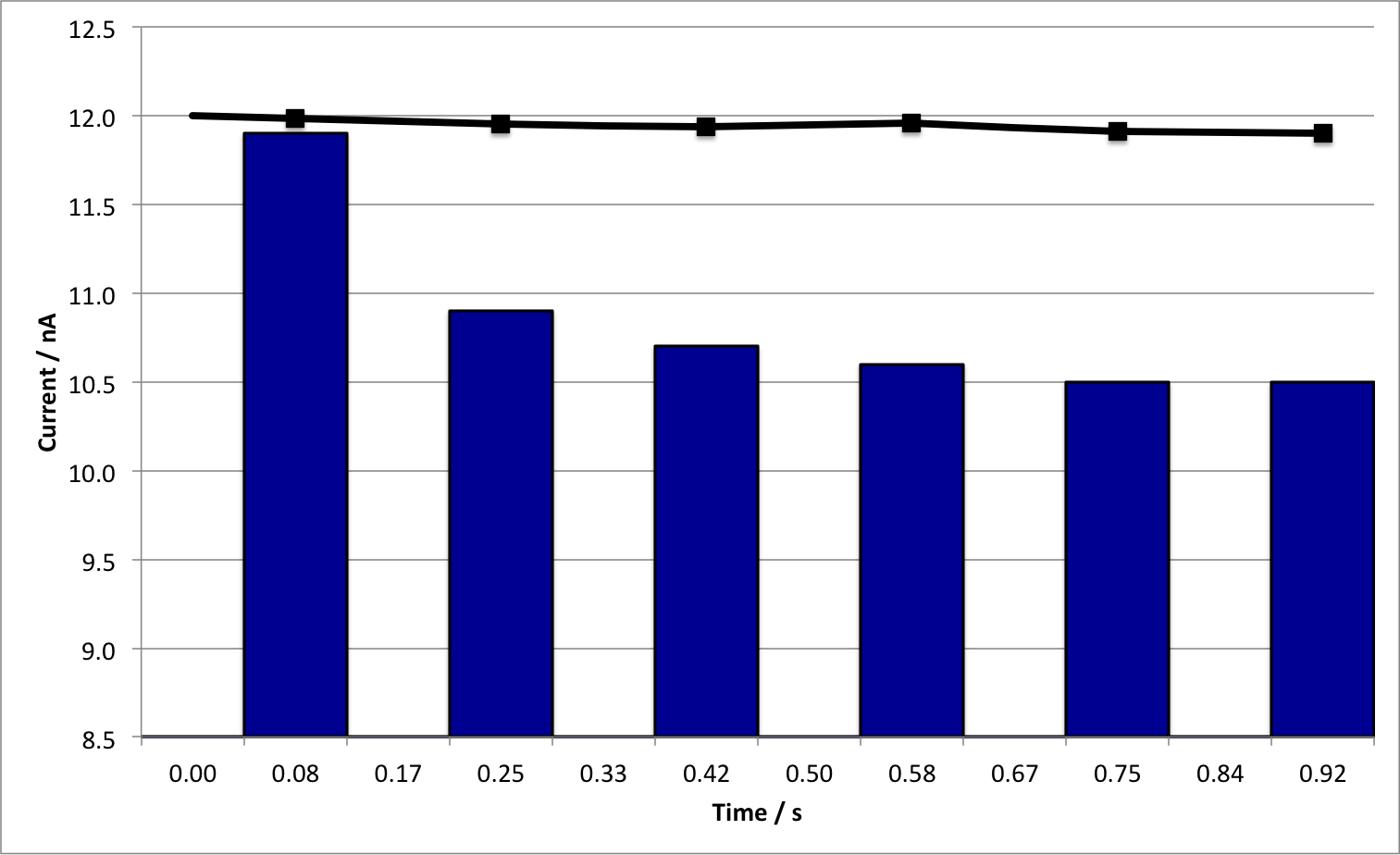}}\\
\subfigure[][]{%
\label{fig:Response}%
 \includegraphics[bb=0 0 576 432,scale=0.3,keepaspectratio=true]{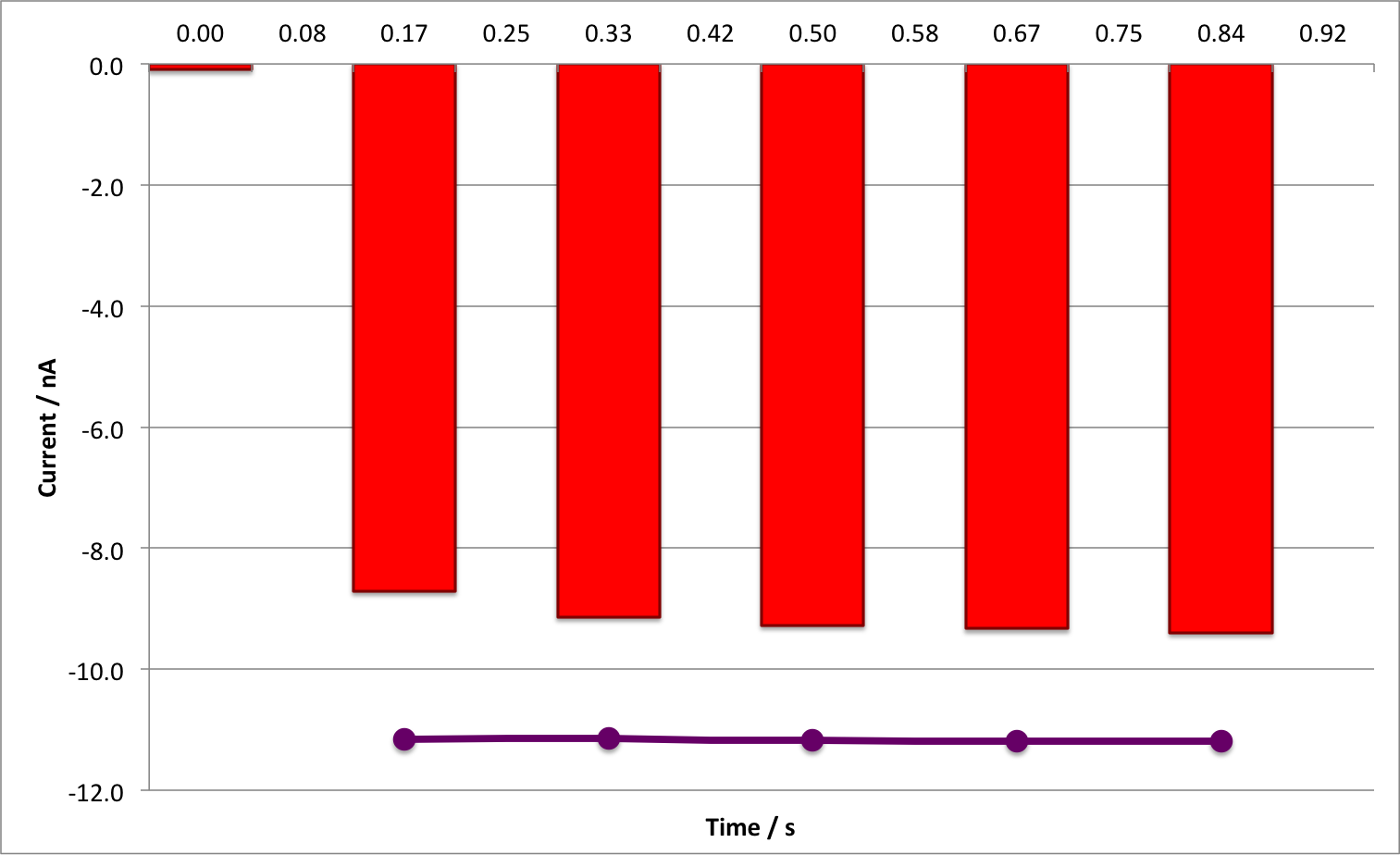}}%
\hspace{8pt}%
\subfigure[][]{%
\label{fig:Voltage}%
 \includegraphics[bb=0 0 576 432,scale=0.3,keepaspectratio=true]{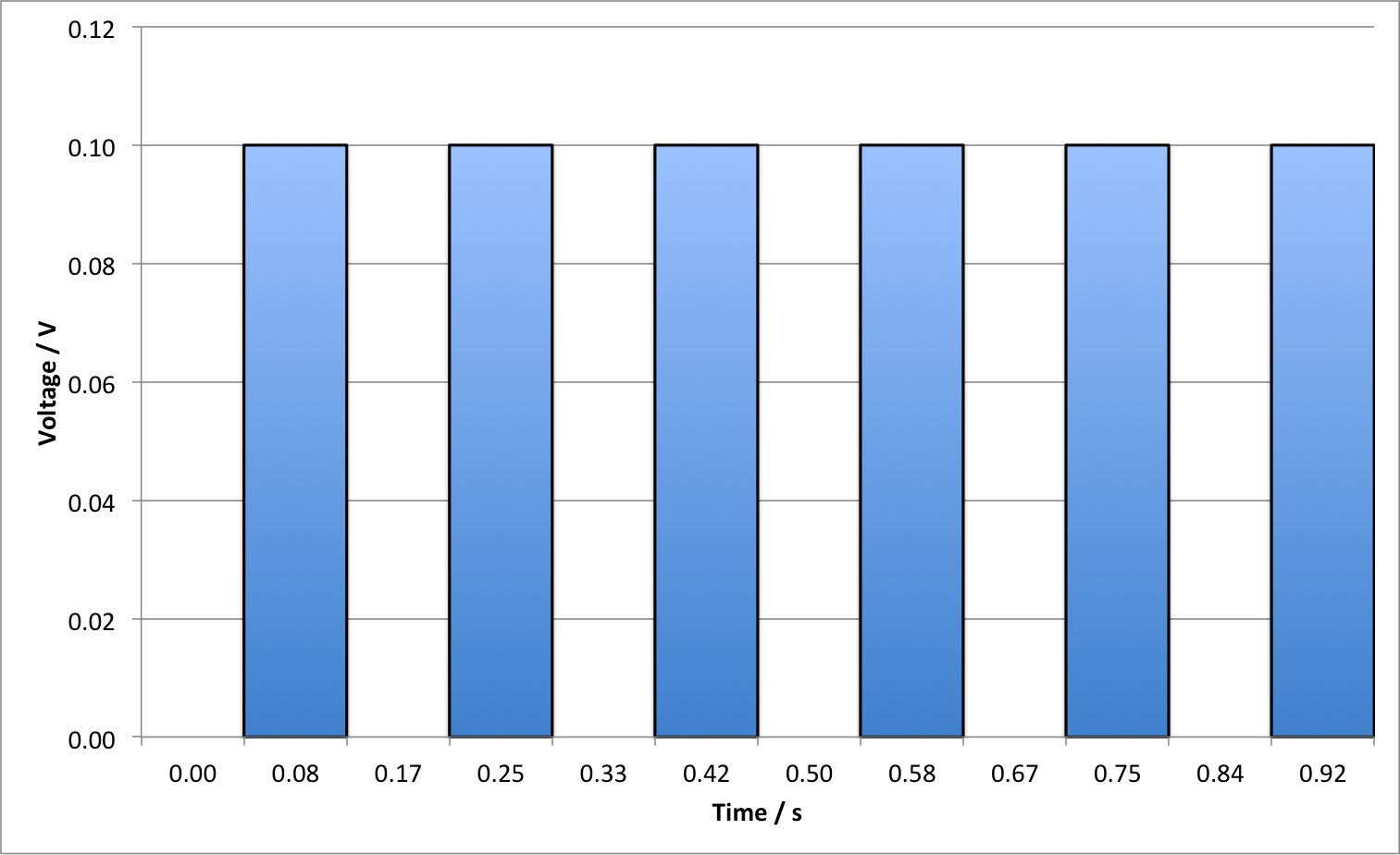}}%
\caption[]{Experimental Replication of the Apalsia experiment:
\subref{fig:Positive} bars: current response to positive spikes applied sequentially; line current response to positive spikes applied after the short-term evaporates;
\subref{fig:Response}  bars: negative `bounceback' response to zeroing voltage, note the first bar is the response to zero after zeroing;
\subref{fig:Voltage} The applied voltage input: 0.1V spikes. N.B. error bars not shown as the electrometer can measure to fA and $\mu$V.}%
\label{fig:ex3}%
\end{figure}

Figure~\ref{fig:ex3} demonstrates the same response in our memristors, part~\ref{fig:Positive} compares the positive current non-habituation response, the line, with the habituation response, the blue bars, to the voltage input in part~\ref{fig:Voltage}. We see a clear decrease in the level of the current response with increasing numbers of stimuli demonstrating electronic habituation with a single memristor. Unlike~\cite{ChuaNanotech} this response is itself a non-linear decrease which follows the same shape as the decay curves in~\cite{243,hystc}. This is (to our knowledge) the first time habituation has been demonstrated in such a manner in a single memristor circuit (although these results were hinted at in~\cite{6815415} which is very interesting as this is a different type of memrisor and suggests that this is a general property of the devices rather than being specific to a material). Unlike neurons, this habituation learning is plastic and only persists as long as the short-term memory of the memristor, allowing for plastic habituation and learning in memristor networks (again, see Erokhin's et al work's for interesting work involving the plastic (rewriteable) and permanent learning in memristor networks~\cite{C2JM35064E}). Our devices also express the state of their memory when the device is returned to zero voltage (bounceback, see section~\ref{sec:rules} for a description) and this is shown in figure~\ref{fig:ex3}\ref{fig:Response}, where the line shows the response of the device to switching to zero after the memory has been lost (non-habituation) and the red bars show the response to zeroing after continued stimuli (habituation), the first bar is the response to the first zero, i.e. after the device has been fully zeroed, and this small value serves as a control for the effect. We see that the negative `bounceback' response (fig 1.b) shows a similar learning effect which is smaller than the positive effect, unlike the positive spike response this effect moves towards the non-habituation response rather than away from it.

\subsection{Experimental Schemes for computing with Memristors}

The result presented in section~\ref{sec:Bio} demonstrate that the memristor is capable of subtraction and we can use this effect to design logic gates.

\begin{figure}[htbp!]
 \centering
 \includegraphics[bb=0 0 576 432,scale=0.35,keepaspectratio=true]{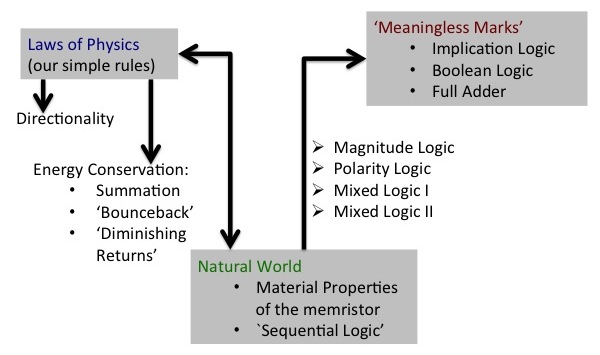}
 \caption{Schematic of the Design of Logical Systems for Computation wiht a memristor}
 \label{fig:Slide}
\end{figure}

If, as suggested by Hilbert, `Mathematics is a game played according to simple rules with meaningless marks on paper', then we can describe computer logic as a game played with the laws of physics in the natural world \footnote{by which I mean the experimentally testable world of the laboratory}. We are free to choose arbitrary properties of the physical world to represent our logic (the `meaningless marks' of the computation game), however, whatever we choose must make sense within the device and how it actually acts (one of the reasons why electronic engineers are so keen on figures of merit that accurately describe what a device will do, rather than theoretical models that tend to be easier to describe and reason about but are not completely accurate). Figure~\ref{fig:Slide} shows the schematic for this process, from experiments and what we know of physical reality we can expect the forward flow of time (which causes directionality, see section~\ref{sec:rules}) and energy conservation to arise from the material properties of the memristor. Directionality suggests the use of a sequential logic approach. By choosing which material property is associated with our computation.

\subsection{Short Term Memory~\label{sec:repro}}

When there is a change in voltage, $\Delta V$, across a memristor the device exhibits a current spike, the physical cause of which is discussed at length in~\cite{243,SpcJ}. This spike is highly reproducible and repeatable and is related to the size of the voltage change ($\Delta V$)~\cite{SpcJ}. The spike's size (as measured by the first measurement after the Keithley's changed voltage) is highly reproducible, the current then relaxes to a stable long-term value (this value is predictable and reproducible), and it takes approximately 2-3 seconds to get to this value. 

This slow relaxation is thought to be the d.c. response of the memristor~\cite{ICNAAM}~\footnote{Note that there has been some discussion as to whether the memristors have a d.c. response [Chua's keynote], within which is has been generally agreed that the perfect memristor may not, but non-ideal memristors (which all real devices are) may. This is not a settled issue and generated much discussion at a recent conference (CASFEST 2014).} and if a second voltage change happens within this time frame, its resulting current spike is different to that expected from the $\Delta V$ alone. The size and direction of this current spike depends on the direction of $\Delta V$, the magnitude of $\Delta V$ and the short-term memory of the memristor. 
 
\begin{figure}[htbp!]
 \centering
 \includegraphics[bb=0 0 576 432,scale=0.5,keepaspectratio=true]{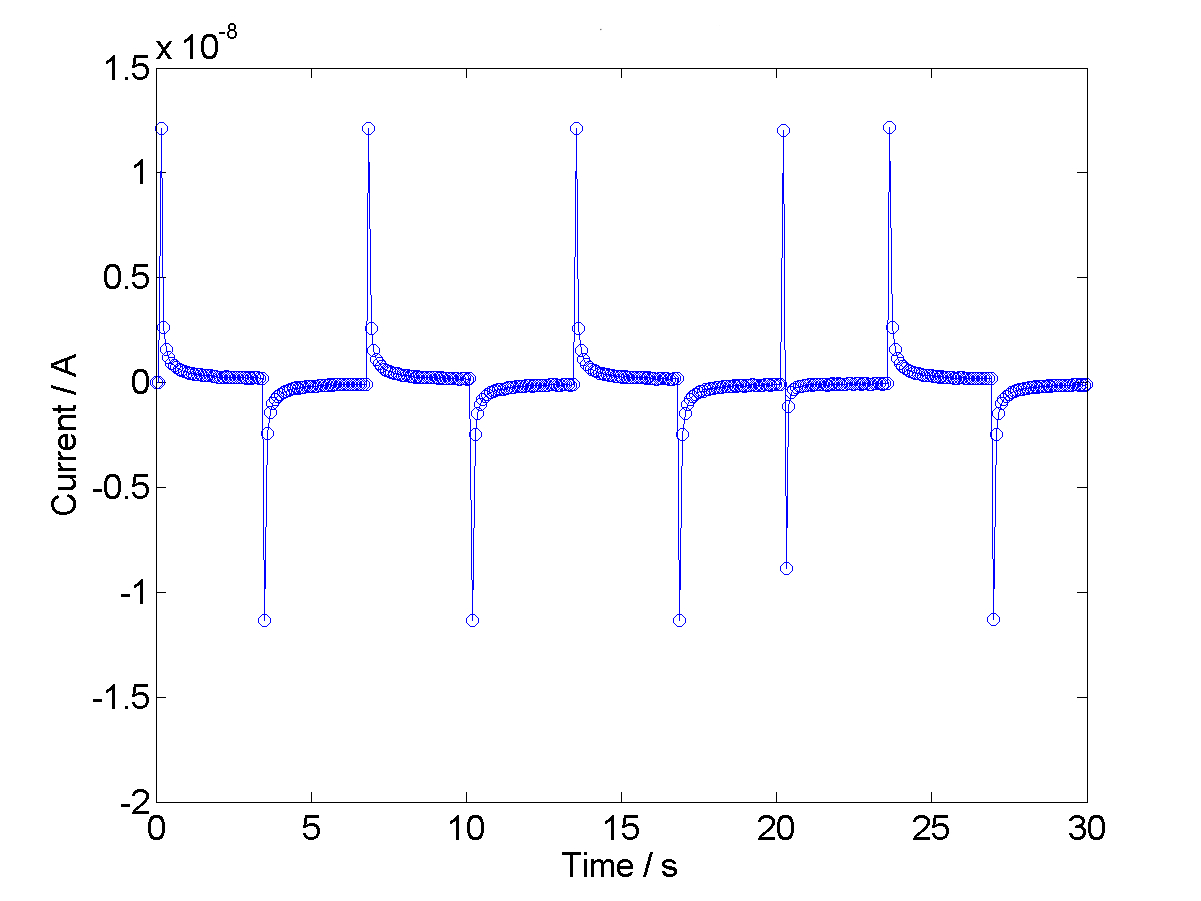}
 \caption{The effect of adding spikes close in time. The response spikes are the negative current spikes. When a positive spike it included but not allowed to relax the corresponding negative spike is smaller.}
 \label{fig:Test1Cu}
\end{figure}

\begin{figure}[htbp!]
 \centering
 \includegraphics[bb=0 0 576 432,scale=0.5,keepaspectratio=true]{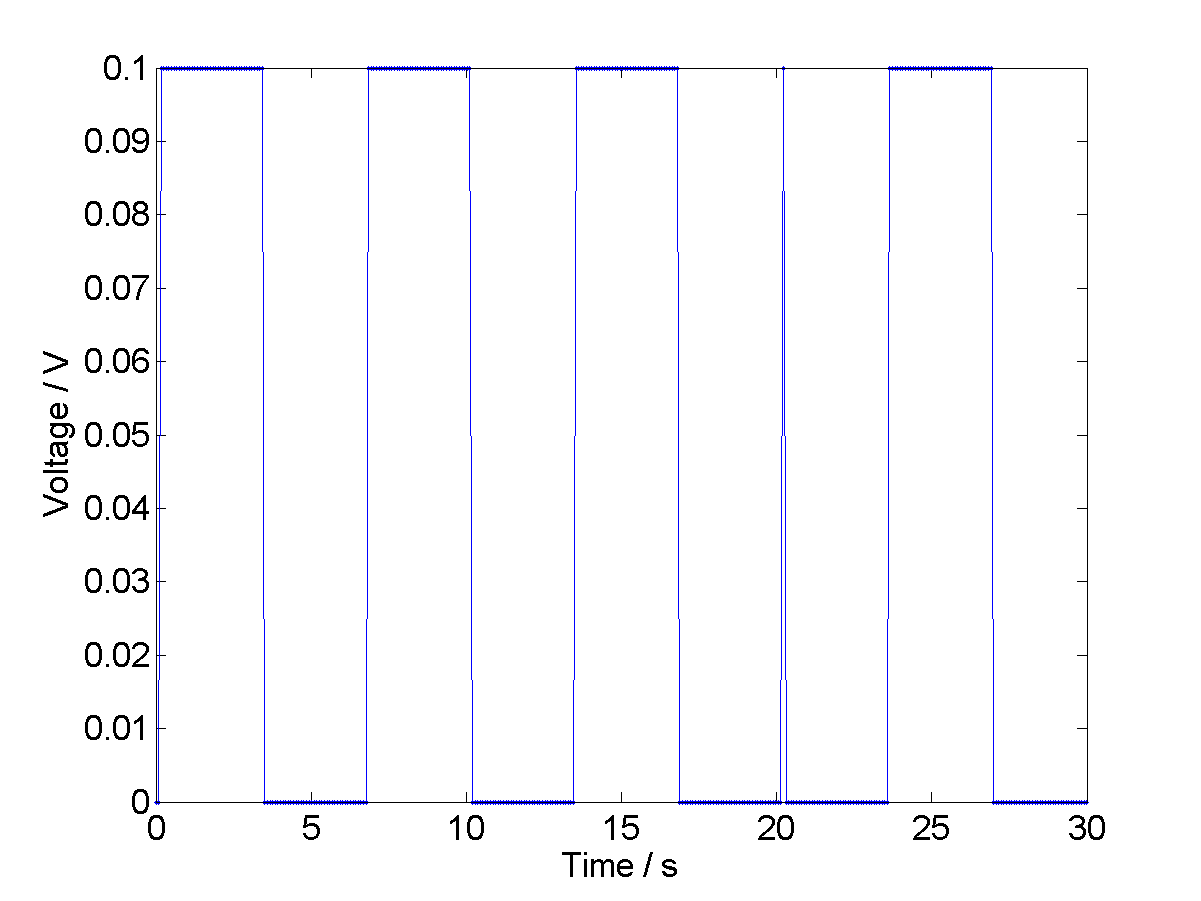}
 \caption{The input voltage for figure~\ref{fig:Test1Cu}}
 \label{fig:Test1V}
\end{figure}

As an example, consider a memristor pulsed with a positive 1V voltage square wave as in figure~\ref{fig:Test1V} (where the pulses are repeated to demonstrate the repeatability) with a timestep of $\approx 0.02s$. The current response is shown in figure~\ref{fig:Test1Cu} and we can see there is a positive current spike associated with the $+\Delta V$ and a perhaps less obvious negative current spike associated with the $-\Delta V$ transition from $+1V \rightarrow 0V$. At approximately 20s, we shortened the square wave to a single time step, and the memory of the system has caused the response spike (responding to the $-\Delta V$ to be smaller (and as it is smaller, it suggests that there is some physical property of the device which has not adjusted to its $+V$ value. See~\cite{Gale} for an discussion on why this physical property is predicted to be the oxygen vacancies in the TiO$_2$.) Thus the response is subtractive in current and additive in resistance state.

\begin{figure}[htbp!]
 \centering
 \includegraphics[bb=0 0 576 432,scale=0.4,keepaspectratio=true]{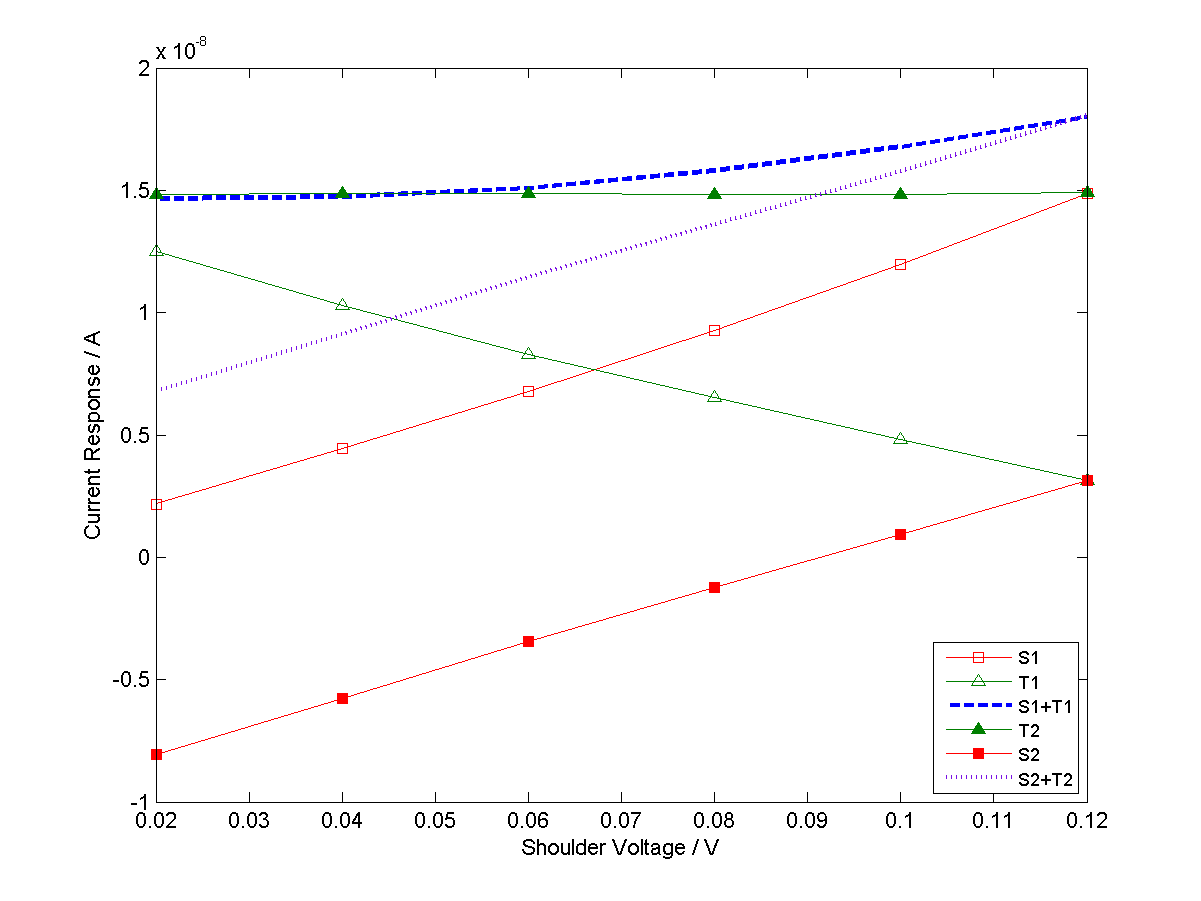}
 \caption{The effects of the order the spikes are sent in to show that spike addition is non-commutative. S1=$\Delta i_{0 \rightarrow A}(t)$, T1=$i_{ A \rightarrow B}(t+1)$, T2=$\Delta i_{0 \rightarrow B}(t)$ and S2=$\Delta i_{B \rightarrow A}(t+1)$. S1 and T1 refer to the shoulder (S1) and peak (T1) currents resulting switching from $0V\rightarrow V_A \rightarrow V_B$. T2 and S2 refer to the peak (T2) and shoulder (S2) of switching from $0V\rightarrow V_B \rightarrow V_A$. In both cases $V_B > V_A$.}
 \label{fig:test10}
\end{figure}

To try and understand the subtleties of this apparent `addition', consider the following system: two voltages, $V_A$ and $V_B$ are sent to the memristor, one after the other separated by one time-step (i.e. before the memristor has equilibrated), where $V_B > V_A$ and $V_B = 0.12V$, and figure~\ref{fig:test10} shows the size of the two resulting spikes as a function of increasing $V_A$. We look at two situations: 
\begin{enumerate}
	\item $V_A(t) \rightarrow V_B(t+1)$; 
	\item $V_B(t) \rightarrow V_A(t+1)$.
\end{enumerate} 
These two situations are drastically different if we look at the transitions, $\Delta V$, as situation 2 has a negative $\Delta V_{B \rightarrow A}$, all the other transitions are positive see table~\ref{tab:steps}. Situation 1 shows that if the smaller voltage is sent first ($V_A \rightarrow V_B$), the current of the first transition $\Delta i_{0 \rightarrow A}$ increases with the size of $V_A$, and the second transition $\Delta i_{A \rightarrow B}$ decreases with the size of $\Delta V_A$, due to the decrease in the effective $\Delta V_{A \rightarrow B}$. However, the sum of these two effects is non-linear, so that the total current transferred (approximated as the sum of the spikes here, but actually the area under the two current transients) is not the same as that shown for situation 2 (until $V_B=V_A$). This shows that more current is being transferred and demonstrates that the spikes are dependent on $\Delta V$. Furthermore, it makes it clear that $\Delta i_{0 \rightarrow A} + \Delta i_{A \rightarrow B} \neq \Delta i_{0 \rightarrow B} + \Delta i_{B \rightarrow A}$, (except in the trivial case where $V_B = V_A$) and that spike based `addition' is non-commutative and therefore the order in which the spikes are sent is relevant to the output of the calculation.  

\begin{table}
\begin{tabular}{|c|c|c|c|}
\hline
$t_1$ 	& $t_2$	& Direction of & $\Delta V$ \\
	&	& energy decrease &  \\
\hline
$V_A$ & $V_A$		& $V_A \goesto V_A$ & $+\Delta V$\\
$V_A$ & $V_B$		& $V_A \goesto V_B$ & $+\Delta V$\\
$V_B$ & $V_A$		& $V_B \leftarrow V_A$ & $-\Delta V$\\
$V_B$ & $V_B$		& $V_B \goesto V_B$ & $+\Delta V$\\
\hline
\end{tabular}
\label{tab:steps}
\caption{Voltage steps for voltage input combinations}
\end{table}

\subsection{Methods of Performing Logical Operations\label{sec:logicmethods}}

\subsubsection{Sequential Logic}

We shall make use of sequential logic (as first implemented in~\cite{P0c}), which works with  the spike interactions seen in the memristor. Memristor sequential logic allows the computation through time by storing a state and allowing it to interact with the input; thus a one terminal device can do two-input (or higher) logical operations, if we are willing to wait for the output. 

\begin{figure}
\centering
\includegraphics[width=2.5in]{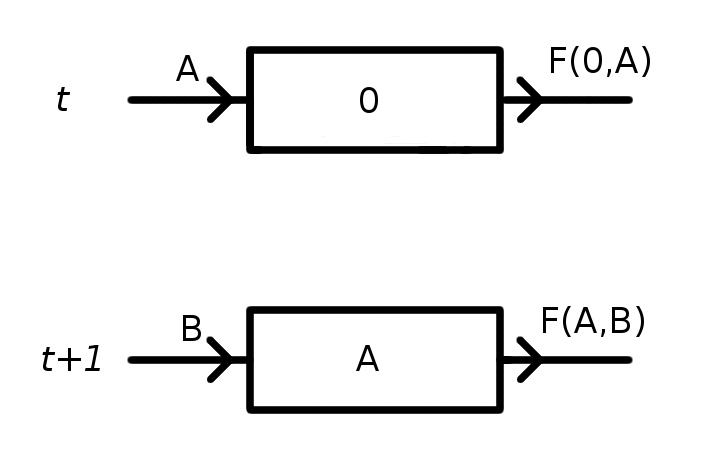}
\caption{Sequential Logic. The output of the memristor is a function of its state, as shown in the box, and the input. As the state is stored for the duration of the short-term memory logical values can be combined if they are input sequentially.}
\label{fig:SeqLogic}
\end{figure}

As shown in figure~\ref{fig:SeqLogic} the memristor's state is stored in its short-term memory and the current output to a voltage change is actually a function of its zeroed/null state and the input. Sequential logic makes use of the memristor's short-term memory to store the first bit, $A$, of an operation before the transmission of the second bit $B$. The output at time, $t_A$, is a function of $A$ and the memristor's starting state (which is $\emptyset$ if the device has been properly zeroed), given by $f(A,\emptyset)$. At time $t_B$ (where $t_B$ is one measurement step after $t_A$) the output would be $f(B,A)$. The response step, $t_1$ is measured one measurement step after $t_b$. Thus, this voltage data is input at $t_A$ and $t_B$ and measured at $t_1, t_2$ and so on where $t_a < t_b < t_1 < t_2$. 

\subsubsection{Elucidated Rules\label{sec:rules}}

Essentially, memristors operate a sequence-sensitive `subtractive summation': where inputs add additional energy to the system, but where that additional energy is subject to subtraction due to energy loss and the whole process operates under nonlinear dynamics. The rules required to design a memristor logic gate follow.

\paragraph{Directionality}

The memristor naturally implements IMP. The memristor is directional: e.g. The response at $t_1$, for $A \rightarrow B$ does not equal the output ($t_1$) for $B \rightarrow A$. The cause of this is that the memristor responds to the difference in voltage. This naturally allows memristor-based sequential logic to compute implication logic as Implication, IMP or $\rightarrow$, requires that $0 \rightarrow 1 \neq 1 \rightarrow 0$ and thus the order in which the two values are input has a meaning. Naturally, sequential logic, as it separates the values in time, implements this ordering. 

\subsubsection{`Summation' via Energy Conservation}

If the logical $\1$ is taken as being a high voltage, i.e. $M$ instead of $m$, then more energy is imparted to the system from the logical combination [$\1,\1$] compared to [$\0,\0$]. This approach can allow the creation of memristor based time-limited summators of use in leaky integrate and fire neurons.

\subsubsection{`Bounceback'}

The application of a voltage spike produces a resultant current spike in the direction of the difference between the starting voltage and the ending voltage, e.g. the first voltage change $V_0 \rightarrow V_A$ causes a positive current response, $+i_A$, if $V_A$ is positive, and negative, $-i_A$, if $V_A$ is negative. If the system is then returned to zero, there is a smaller current spike of the opposite polarity, i.e. $-i_0$ and $+i_0$ respectively for the two examples mentioned above. If several spikes are input before returning to zero, i.e. a sequence of $[V_0, V_A, V_A, V_0]$ the current spike, $i_0$ is larger than would be the case for $[V_0, V_A, V_0]$, although not twice as large due to losses in the system (as the system is nonconservative).

\subsubsection{`Diminishing Returns'}

As discussed in the example of $[ V_0, V_A, V_A, V_A ]$ above each addition of each consecutive spike has a reduced effect compared with the first. This property is seen with spikes of hte same polarity and with changing polarity i.e. the response spike to $[V_0, +V_A, -V_A, +V_A, -V_A]$ is smaller than $[V_0, +V_A, -V_A]$.  This only happens to spikes input into the memristor's short-term memory as waiting for the device to return to a blank state refreshes the property that reacts to the voltage step. We expect that the physical property in question is related ot the ions in the device, as it they that `hold' the memory, but this has not been experimentally verified. The diminishing returns effect is behind the habituation shown in section~\ref{sec:Bio}.

\subsection{Examples of Logical Systems\label{sec:examples}}

Knowledge of these rules and effects allows us to design logical computation systems which perform a surprising amount of computation with only a single memristor. We have found that, in these schemes that the summation effect is important in magnitude logic, the `bounceback' effect is more relevant in polarity logic (although both affect the outcome). As these can be balanced and set in opposition to each other, the richest effects came from using the mixed logics (as presented in~\ref{tab:Logic}: we will now present a few examples.

There are two variables we can utilise when assigning logical values: the magnitude, as represented by $M$ for a high magnitude and $m$ for a low magnitude; and the sign, as represented by a $+$ for positive and $-$ for negative. The 4 different logical assignations that can be applied using these values is shown in table~\ref{tab:Logic}. To implement logical operations, voltage spikes are applied for one time-step and the response recorded at the same frequency. In between logical operations, the devices were left for longer than the equilibration time ($\tau_{\infty}$ in~\cite{SpcJ} which is around 3.5s) to zero the memristor by removing its short term memory. 

%
\begin{table}
\caption{Four different methods of implementing logical $\1$ and $\0$ with memristor spikes: $M$ refers to a high magnitude voltage, $m$ to a low magnitude voltage and `+' and `-' refer to its polarity.}
\label{tab:Logic}
\begin{center}
\begin{tabular}{|c||c|c|c|c|}
\hline
Logical 	& Magnitude	& Polarity	& Mixed 	& Mixed 	\\
value		& Logic		& Logic& Logic 1	& Logic 2	\\
\hline
One 		& M 			& +		& +M		&  -M	\\
Zero 		& m 			& - 	& -m		& +m\\
\hline
\end{tabular}
\end{center}
\end{table}

Changing the values of $M$ and $m$ can allow the results to be tuned or balanced against the effect of polarity, but in this paper we shall just deal with qualitative examples. From investigation of these systems, we have elucidated the following physical rules for the system.

\subsection{Simple Logic Gates~\label{sec:simplegates}}

We can do Boolean logic with the spike interactions by sending the second bit of information one time-step (0.02s) after the first. We take the input as the current spikes from the voltage level. The output is the response current as measured after the 2$^{\mathrm{nd}}$ bit of information. at that timestep, i.e. $V_B$. After a logic operation the device is zeroed by being taken to 0V for approximately 4s, and this removes the memristor's memory. 

We have some freedom in how we assign the $\1$ and $\0$ states to device properties and these give different logic. The following examples will demonstrate some approaches and build an OR gate or an XOR gate: i.e. $\1$ is any positive current output, $\0$ is any negative current output, inputs are positive ($+1$) and negative ($\0$) voltages.

\subsubsection{Inverter}

\begin{figure}
\centering
\includegraphics[width=3.5in]{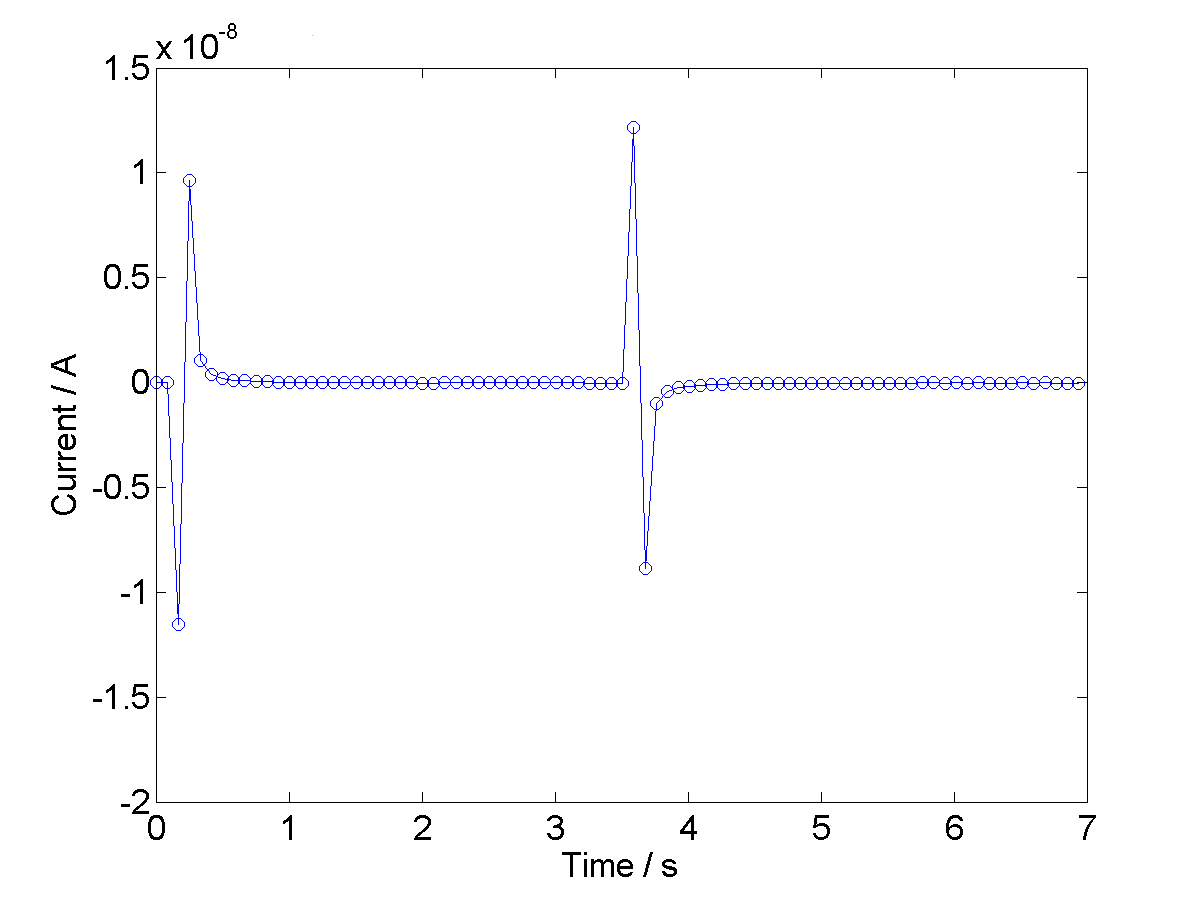}
\caption{AND Gate Implementation}
\label{fig:NOT}
\end{figure}

Using polarity logic and the `bounce-back' effect, it is an easy thing to build an inverter as shown in figure~\ref{fig:NOT}. Because the response spike is in the opposite direction, taking that as the result of the operation switches from $\1$ to $\0$ (and vice versa) and can be viewed as performing the NOT operation on the input.

\subsubsection{AND Gate}

An example of an AND gate is shown in column 5 in figure~\ref{fig:AND}, this example uses mixed logic 2 (see table~\ref{tab:2bitgate}) with a $M$ of -0.5V and a $m$ of +0.001V. If we take the response output as $\1$ if current over a threshold (in this case, 0.55$\mu A$) is seen, the device implements an AND gate (this is still the case if we choose to limit ourselves to only the value of the $t_1$ response spike). Due to the summation effect, the amount of energy in the [$\1,\1$] system is larger than the [$\0,\1$], [$\1,\0$] and [$\0,\0$] parts of the truth-table, and this causes a larger `bounceback' response which can be measured in the positive current response. 

Were we to limit ourselves to the negative current part of the device response, the magnitude of the output picks out an inclusive OR operation, in that the only parts of the truth table that have a response over the threshold are those that contain a $\1$ (because these spikes depend on a $\1$ input). Although this response is trivial, it is information that can be usefully used with the correct output circuitry.

\subsubsection{OR Gate~\label{sec:OR}}


The truth table for an OR gate is given in column 8 in table~\ref{tab:2bitgate}, essentially, the output should be $\1$ if either of the inputs contained a $\1$. We take the $\0$ output as being below a threashold current and the $\1$ output as being above a threashold. The threashold is set to $>$18nA with the $\0$ input being set of 0.01V and the $\1$ as 0.2V~\footnote{Using $\0$ as 0V was also tested, it works and is lower power but was not chosen as an example as it is a trivial case.}, which gives the voltages below:
 
\begin{itemize}
\item $\0$, $\0$ = 0.01V, 0.01V
\item $\0$, $\1$ = 0.01V, 0.2V
\item $\1$, $\0$ = 0.2V, 0.01V
\item $\1$, $\1$ = 0.2V, 0.2V.
\end{itemize}

\begin{figure}[htbp!]
 \centering
 \includegraphics[bb=0 0 576 432,scale=0.5,keepaspectratio=true]{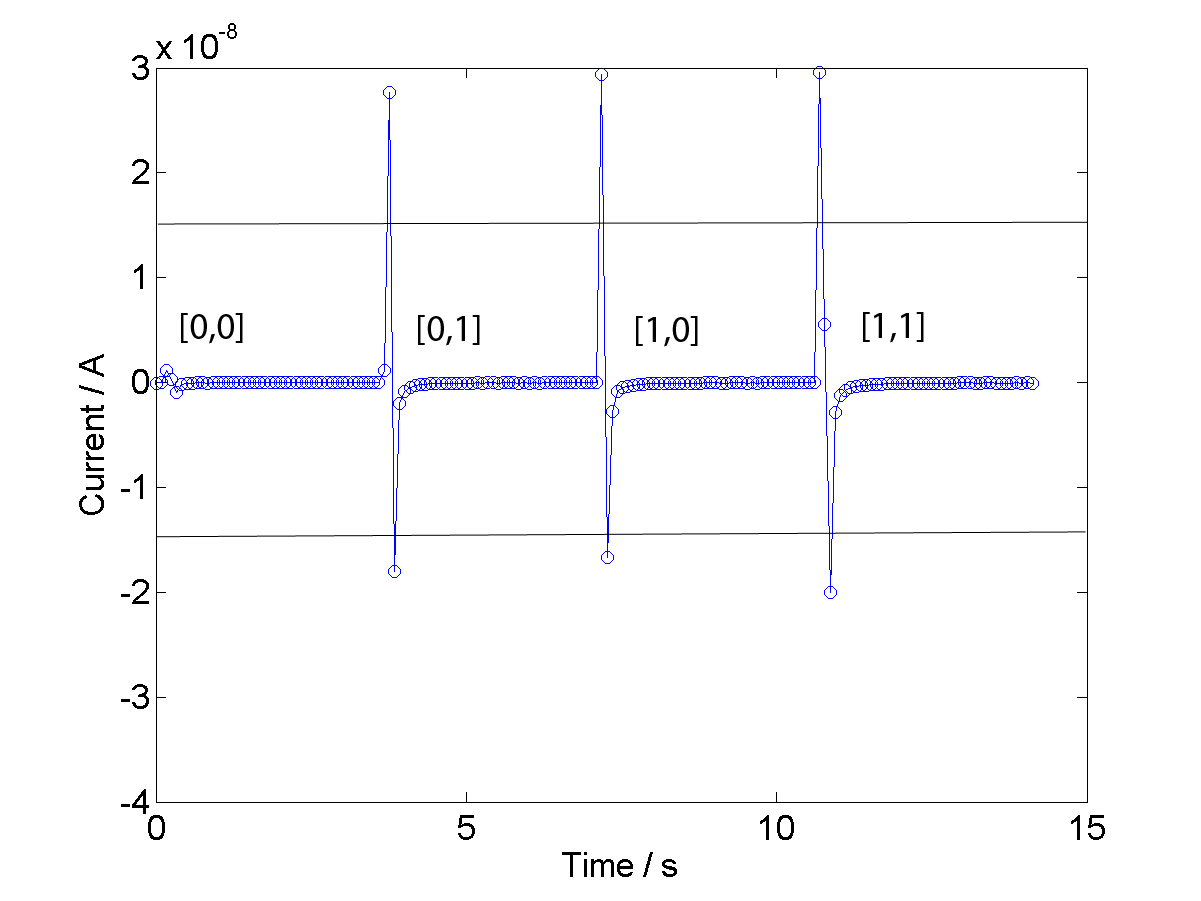}
 \caption{OR Gate. Using $\1$ equal to a current spike caused by a voltage change to 0.2V and $\0$ equal to a current spike caused by a voltage change to 0.01V we can make a serial OR gate (where logical 1 is considered to be a current which is more than 5nA). At 0.04s `0, 0' was input, giving peaks below the threashold i.e. $\0$ as an output. The three large peaks are $\1$ outputs resulting from `0,1',`1,0' and `1,1' inputs.}
 \label{fig:LogicTest3}
\end{figure}

Figure~\ref{fig:LogicTest3} shows the current data from the voltage inputs above. It can be seen that when a $\1$ is input, there is a large spike output. To read the logical state of the device, one merely takes the current value as the second bit is read in. 

\subsection{Complex Logic Gates~\label{sec:XOR}}

The XOR truth table is shown in column 9 in table~\ref{tab:2bitgate}. If we take logical $\1$ to be the current resulting from a positive voltage and a logical $\0$ to be the current resulting from a negative voltage, then, the response is the current when the 2$^{\mathrm{nd}}$ bit is input (not after, although it could be designed that way but it is slower). We get a high absolute value of current if and only if the two inputs are of different signs, i.e. we have $\{ \1\:\0 \}$ or $\{ \0\:\1 \}$ which gives us an exclusive OR operation. For this logical system, we used the same voltage level and allowed a change in sign to indicate logical zero or logical one:
\begin{itemize}
\item $\0, \0$ = -0.1V, -0.1V
\item $\0, \1$ = -0.1V, +0.1V
\item $\1, \0$ = +0.1V, -0.1V
\item $\1, \1$ = +0.1V, +0.1V.
\end{itemize}

\begin{figure}[htbp!]
 \centering
 \includegraphics[bb=0 0 576 432,scale=0.5,keepaspectratio=true]{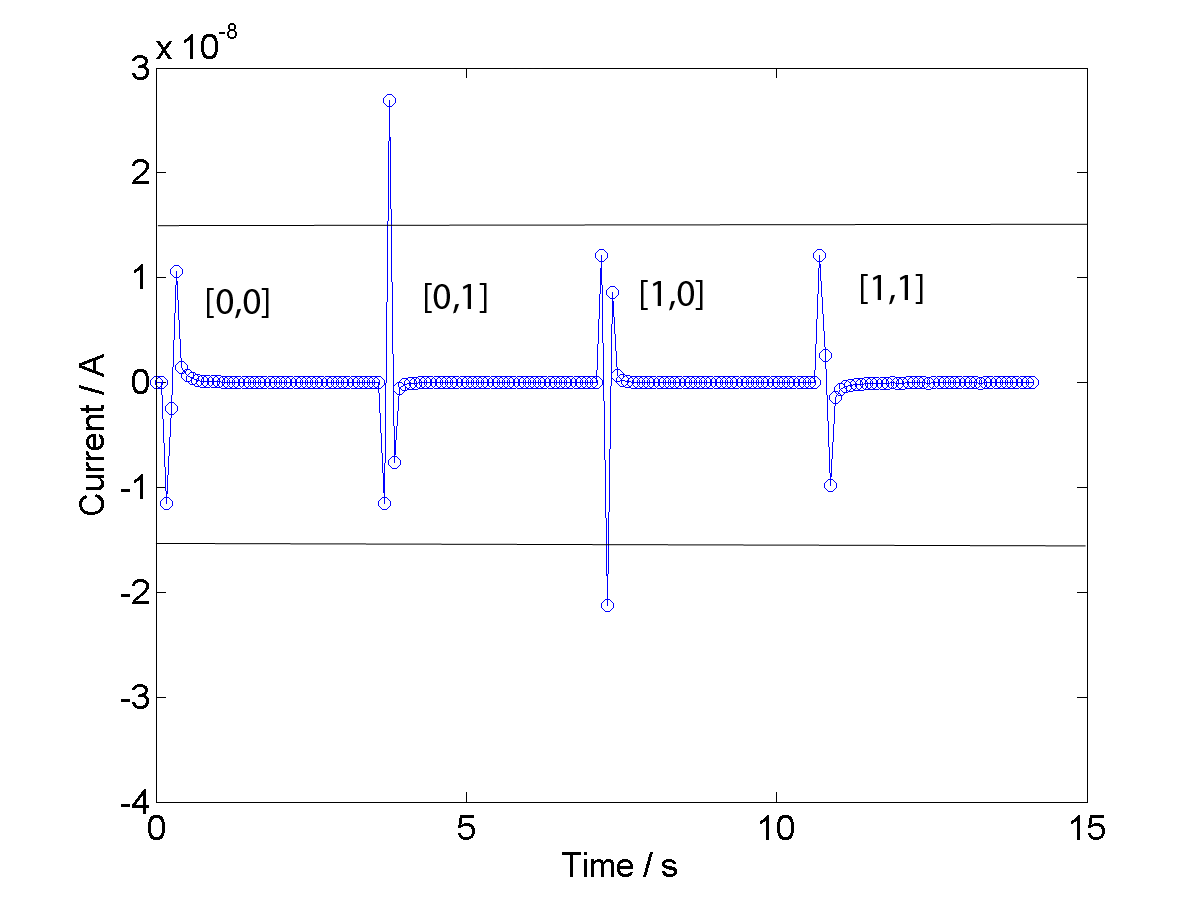}
 \caption{XOR gate, the modulus of a current response over $\pm 1.25\times10^{-8}$A is taken as one, as current response under that threashold is taken as zero.}
 \label{fig:Test7Current}
\end{figure}
\begin{figure}[htbp!]
 \centering
 \includegraphics[bb=0 0 576 432,scale=0.5,keepaspectratio=true]{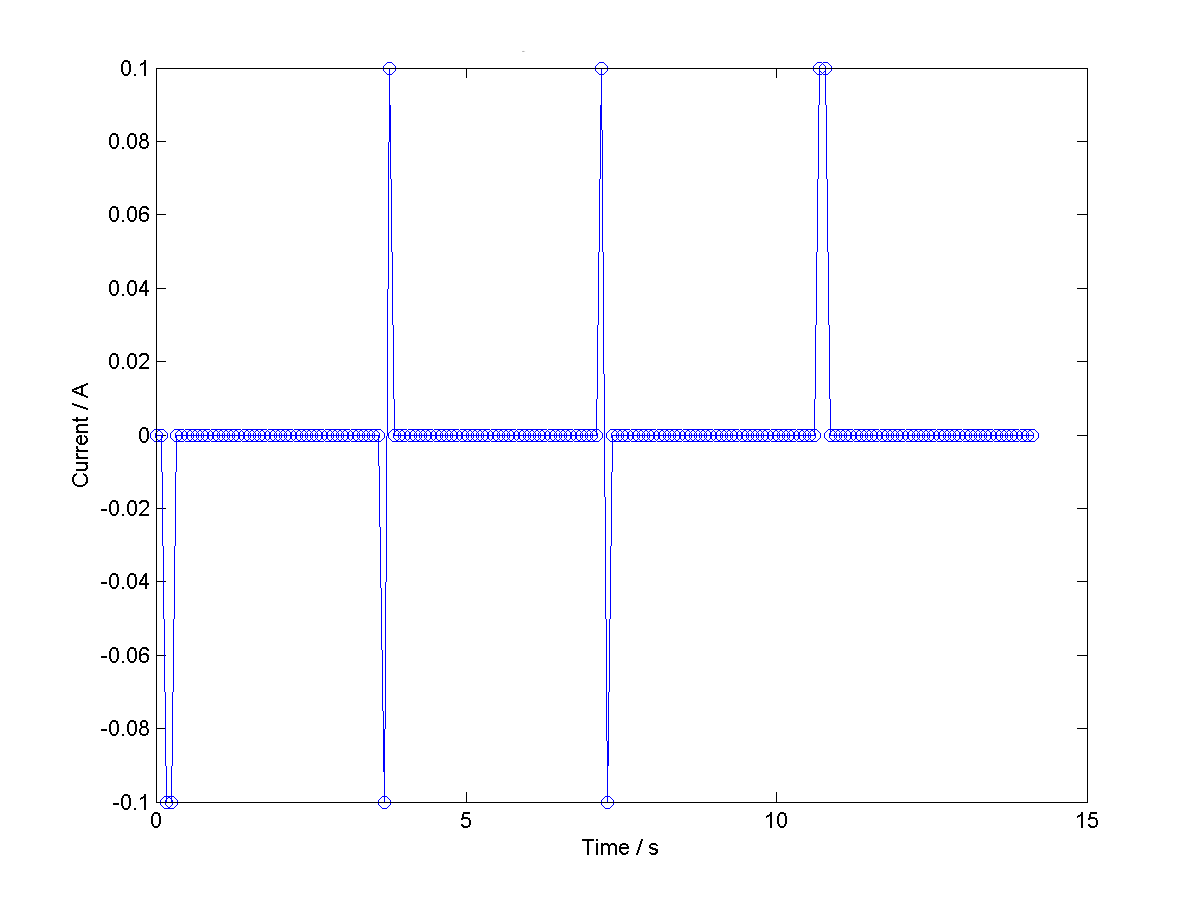}
 \caption{The input voltage for the XOR gate.}
 \label{fig:VoltageXOR}
\end{figure}

As an example, the input voltage is shown in figure~\ref{fig:VoltageXOR} and the current output is shown in figure~\ref{fig:Test7Current}. This is based largely on `bounceback', a $\1$ or $\0$ input to a memristor `holding' the opposite polarity in memory causes a bigger change. Essentially we measure the convolution of the `state' and `input', represented as a function $F[S,A]$ i.e. $F[\0,\1] \approx F[ \1, \0 ] > F [ \1, \1 ] \approx F [ \0 , \0 ]$.

\begin{figure}[htbp]
 \centering
 \includegraphics[scale=0.5,keepaspectratio=true]{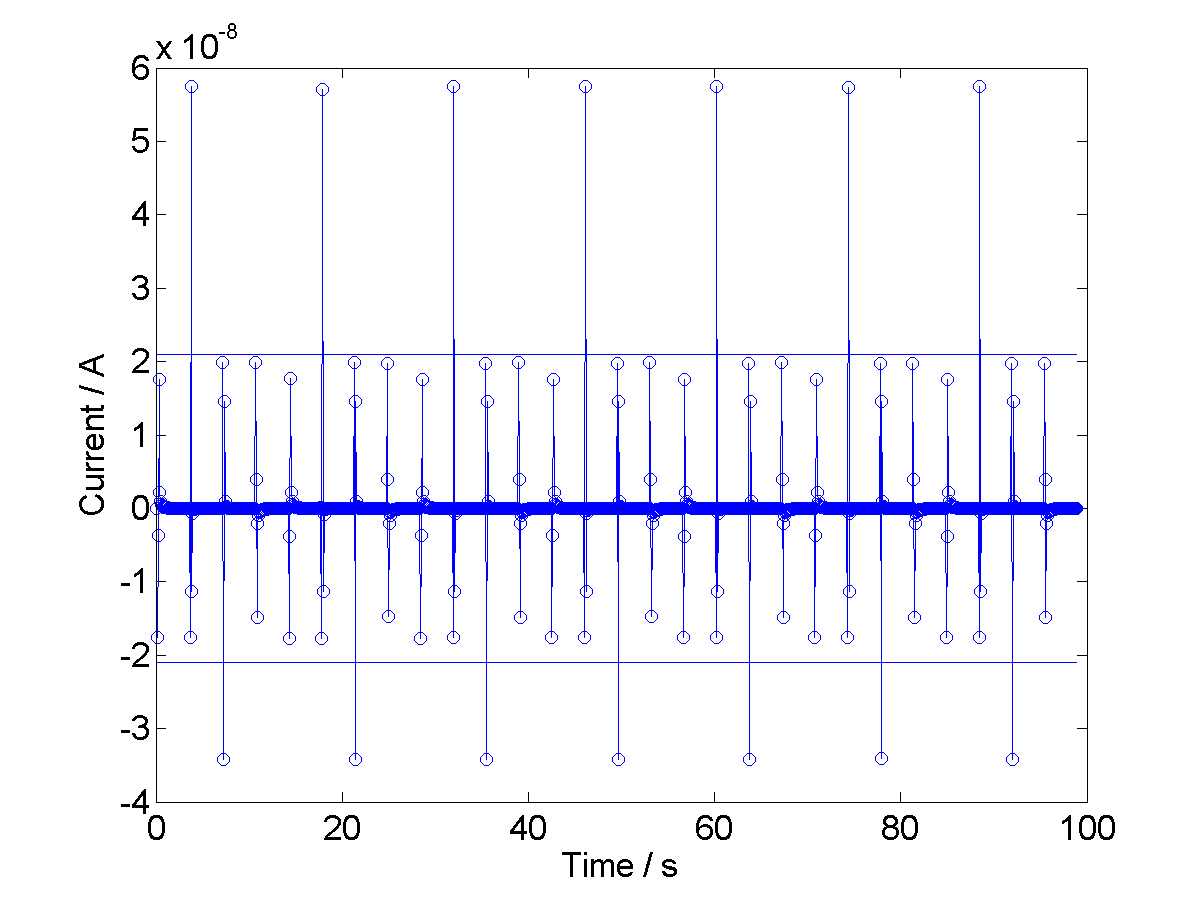}
 \caption{Reproducibility test of XOR function. Here the XOR truth table is run 7 times (using a different set of voltage input values). The threshold between $\1$ and $\0$ is marked as shown.}
\label{fig:XORReproTest}
\end{figure}

With a pause between operations to allow the memristor to lose its memory, the XOR operation is reproducible, as shown in figure~\ref{fig:XORReproTest}. Note, as our devices are slightly asymmetric, $F [ \0, \1 ]$ does not exactly equal $F [ \1 , \0 ] $ and $F[ \0,\0 ]$ does not exactly equal $F [\1 , \1 ]$. This arises fomr the material and the method of synthesis (see~\cite{260}) and has nothing to do with the voltage being postive per se, the devices are always wired up the same way round before measurement. Larger voltage ranges (see figure~\ref{fig:XORReproTest}) highlight the difference between positive and negative, demonstrating how different logical combinations can be split out or alternatively balanced by sensitive voltage choice.

As XOR A = NOT A, if we always take the 2$^{\mathrm{nd}}$ point after the first (and only bit in this case) as being the response bit (as we did above for the XOR gate), we have a NOT gate.

\subsection{Full-Adder~\label{sec:fulladder}}

\begin{figure}
\centering
\includegraphics[width=3.5in]{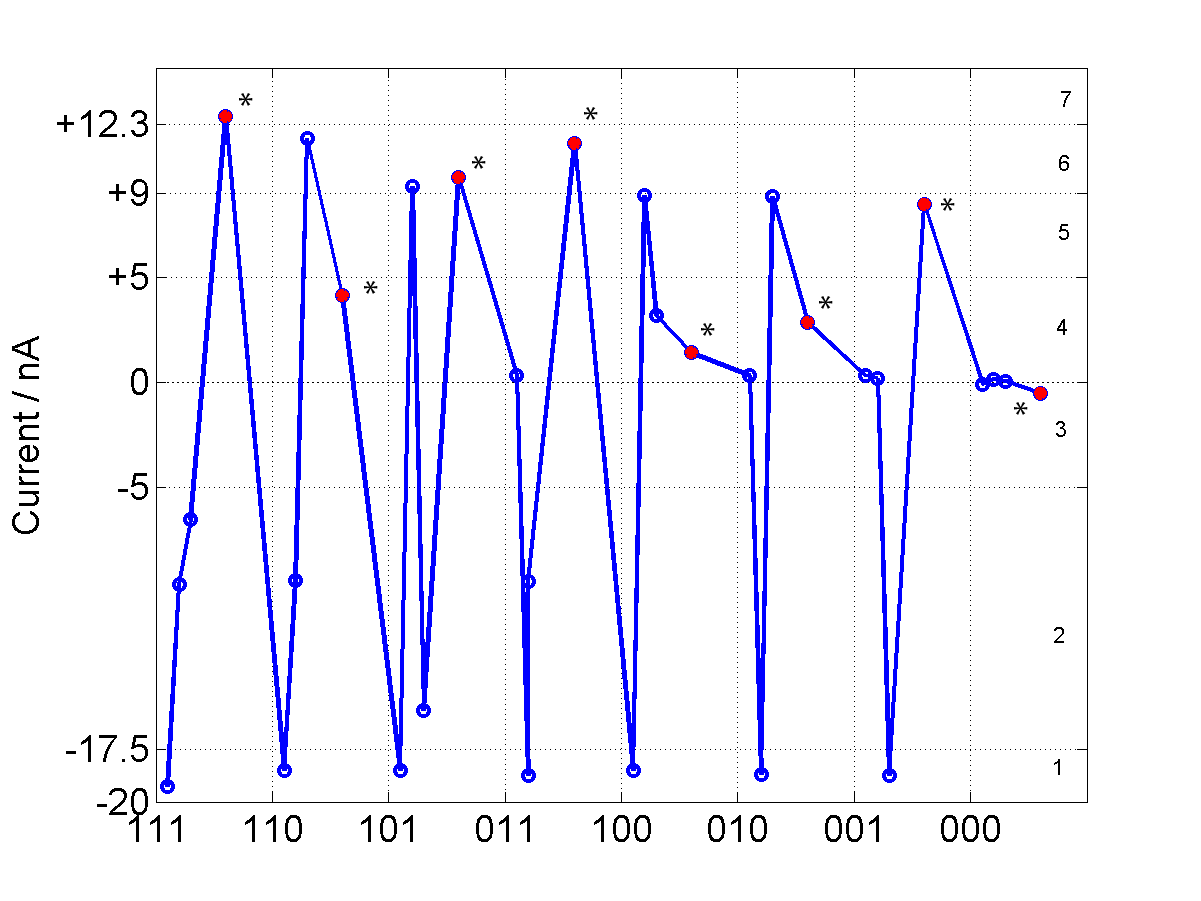}
\caption{A full adder using mixed logic 2. The first three input bits are the logical inputs, the system has one timestep to respond (\~1s) before a read spike is sent in, as marked by an *. The numbers of the ranges correspond to the list.}
\label{fig:FullAdder}
\end{figure}

It is possible to compute an unconventional instantiation of full-adder, as shown in figure~\ref{fig:FullAdder} (admitting that we require a voltage spike to current spike conversion). The two input and carry bits are input as a series of spikes using mixed logic 2 with input $\1$ represented by -0.5V and input $\0$ represented by +0.001V. The input sequence is [A,B,C,1,2,3,4], with the logic input at $t_A-t_C$, the response spike recorded at $t_1$, an extra read voltage of -0.15V input at $t_2$. This gate requires a clock to operate. Figure~\ref{fig:FullAdder} shows the response of the memristor to this scheme, for the three inputs of a full adder, the read spike at $t_2$ is marked with an * to make it easier to understand, and the data of the memristor losing its short-term memory is not shown. 

From this set-up the following things can be deduced from knowing the maximum positive and negative current spikes within 4 time-steps of an input (although this requirement need not be too stringent if we have a way of recording the maximum current within the ranges in between zeroing the system, which we can do with knowledge of the read pulse clock). 

The resulting information from the current is thus:
\begin{enumerate}
\item if a negative current is recorded in the range -17.5 to -20nA: we have had a $\1$ input into the system
\item if a negative current is recorded in the range -5 to -17.5nA: we have a carry bit from the operation
\item if a negative current is recorded in the range 0 to -5nA: we have had a zero in the system (this is redundant information)
\item if the maximum positive current is recorded in the range 0 to +5nA: the result of the calculation is $\0$
\item if the maximum positive current is recorded in the range +5 to +9nA: the result of the calculation is $\1$
\item if the maximum positive current is recorded in the range +9 to +12.3: the result is `2' (or $\1$ for the carry bit, $\0$ for the summation bit) 
\item if the maximum positive current is recorded over 12.5nA: the result is `3' (or $\1$ for both the carry and summation bit in binary logic). 
\end{enumerate}

The output in the negative is purely a result of the input voltages to the system. The positive system includes the `bounceback', and the summation effect as probed by the read voltage which gives threasholded values of the memristor's state. 

With switches, it would be possible to send on the logical result as binary. Region two of the plot encodes the carry bit for the operation, because only if there are two $-M$ spikes (which encode $\1$) within 3 time-steps of each other we will see a current response in that range. The summation bit is not encoded in as direct a manner, the maximum of the positive currents encodes the numerical sum, and so the summation bit for the value 3 is in a different place to that for the value 1. If we only require knowledge of the carry and summation bit, we can do without the read voltage and corresponding spikes. Changing the values of $M$ and $m$ can tune the effect and might allow us to change the relative values of the output spikes.

\begin{table*}[t]
  \centering
  \begin{tabular}{|cc||ccccccccc|ccccccc}
    \hline
    $p$ & $q$ & 1 & 2 & 3 & 4 & 5 & 6 & 7 & 8 & 9\\
    \hline
    \hline
    $\0$ & $\0$ & $\1$ & $\0$ & $\1$ & $\1$ & $\1$ & $\0$ & $\1$ & $\1$ & $\0$ \\
    $\0$ & $\1$ & $\0$ & $\0$ & $\1$ & $\0$ & $\0$ & $\1$ & $\1$ &  $\1$ & $\1$ \\
    $\1$ & $\0$ & $\0$ & $\0$ & $\0$ & $\1$ & $\0$ & $\1$ & $\1$ &  $\0$ & $\1$ \\
    $\1$ & $\1$ & $\0$ & $\1$ & $\0$ & $\0$ & $\1$ & $\0$ & $\0$ &  $\1$ & $\1$ \\
    \hline
    \multicolumn{2}{|c||}{Name} & NOR & AND & Not p & Not q & NXOR & XOR & NAND & IMP & OR \\
    \multicolumn{2}{|c||}{Symbol} & $\NOR$ & $\+$ & $\Not p$ & $\Not q$ & $\NXOR$ & $\XOR$ & $\NAND$ & $\IMP$ & $\Or$ \\
    \multicolumn{2}{|c||}{No.*}& $F_8$ & $F_1$ & $F_{12}$ &  $F_{10}$ & $F_9$ & $F_6$ & $F_{14}$ & $F_{13}$ & $F_{7}$ \\
    \hline
  \end{tabular}
  \caption{Selected 2-bit binary Boolean functions and their corresponding gate names (as used in this work). *Function number from~\cite{376}.}
  \label{tab:2bitgate}
\end{table*}

\subsection{Logical Complexity of a Full Adder\label{sec:LCFA}}

A half adder is a 2-bit output basic gate, it is equivalent to an XOR and an AND gate. 

Now we are in a position to analyse our full adder from the point of view of logical efficiency. The 2-bit full adder separates 8 3-bit inputs into 4 arithmetical groups. arbitrarily large numbers can be added by chaining full adders as the carry bit out to the carry bit ($C$) in. Using the same description as before, as standard full adder has $N=8$ ($2^3$) possible inputs, and 4 distinguishable operations, giving us a logicial efficiency of 50\%. Our spiking logic full adder allows the differentiation of all 8 operations, and gives us a logical efficiency of 100\%. Using standard circuit complexity measures of circuit depth complexity and circuit size complexity we get 1 + selection circuitry for our full adder and [x] for a standard transistor based full adder. This is not a fair comparison, of course, because a standard full adder has multiple ports and all bits are received at hte same time, in our full adder we have to take 2 timesteps to get the answer.

As an aside, this measure of logical efficiency means that the full adder is less logically efficient than a half adder. A half adder requires 2 input bits and 2 output bits ($B^{2} \goesto B^2$) and can separate out 4 inputs into 3 groups, giving a logical efficiency f 75\%. The full adder has to add an entire extra bit to count up to 3 and separate out 8 inputs into 4 groups. Of course, a single logic gate doing a full adder's truth table is better (as it involves 3 bits of input not 2*2=4) and of course he full adder is more useful. However the half adder is 75\% efficient on 4 inputs, allowing it to separate out 3 different inputs, the full adder is 50\% efficient on 8 inputs which allow it to separate out 4 groups of inputs.

\begin{table}
\centering
\begin{tabular}{|ccc|cc|c|}
\hline
$A$ 	& $B$ 	& $C$ 	& $\Sigma$ 	& $C$	& Arithmetical$\Sigma$ \\
\hline
$\0$	& $\0$	& $\0$	& $\0$			& $\0$	& 0 \\
\hline
$\1$	& $\0$	& $\0$	& $\1$			& $\0$	& 1 \\
$\0$	& $\1$	& $\0$	& $\1$			& $\0$	& 1 \\
$\0$	& $\0$	& $\1$	& $\1$			& $\0$	& 1 \\
\hline
$\1$	& $\1$	& $\0$	& $\0$			& $\1$	& 2 \\
$\1$	& $\0$	& $\1$	& $\0$			& $\1$	& 2 \\
$\0$	& $\1$	& $\1$	& $\0$			& $\1$	& 2 \\
\hline
$\1$	& $\1$	& $\1$	& $\1$			& $\1$	& 3 \\
\hline
\end{tabular}
\caption{Full adder truth table}
\label{tab:FA}
\end{table}


Relatively simply, we can immediately write that for each input port removed from the gate, an extra input time-step must be added. Thus, we can replace circuit connections with extra time steps. Thus we arrive at an answer to the `convesion rate' between space and time where we count space by the number of input wires and time by the time-step (in a spiking clocked system). It seems to be 1:1. In standard electronics a full adder involves 3 input wires (one time-step), in the memristor system, we use 1 input wire and 3 time-steps. Output in standard electronics is 2 output wires, in the memristor system we can either use 1 time output plus.

\section{Conclusions~\label{sec:conc}}

In this paper I have shown: the use of asyncronous logic based on spikes, elucidated the rules governing spike interaction, invented logical assignations for these physical processes, used them to design gates and demonstrated a full adder that possesses extra functionality that leads to it being a logically-reversible gate. This was all done with a single memristor plus the additional circuitry to convert current output spikes to voltage inputs (but we have no reason to expect that this will be bigger than the current state of the art). 

These results movtivate the question of how far can we take this approach. I have no reason to believe that a 2-bit full adder is the limit of the functionality of a memristor (although I posit that we can't take it too much further with our devices without reaching a detection limit). In fact, using the memristor as a synapse, integrator or habituator could be considered more complex than a 2-bit full adder.

This paper also suggested a measure of efficiency to compare (computation) time and (circuit) space complexity, and this measure was used to compare the spiking memristor full adder with a standard full adder and put a number to that improvement.

Nonetheless, this method of full adder creation has possible drawbacks in that one has to wait longer for the solution to a posed question. This is less relevant than in might seem for some uses. The brain is much slower than conventional electronics. Furthermore there is the idea that Moore's law is coming to an end, and the current request from~\cite{322} is the concept of `More-than-Moore', where a device has extra functionality `riding on top of' standard shrinkable hardware. 

This work shows that by using the memristor's extra functionality of possessing a memory, we can do a seemingly amazing amount of computation. There are some drawbacks, of course. Our memristors are slow and most memristors operation on neural-spike-like speeds, which are far slower than modern day circuits. To use these logic gates, we need to convert current spikes back into voltage spikes, the circuits to do this are relatively trivial (op-amp, or a 2:1 stage), but will take up circuit board area, so the size reduction isn't quite as good as calculated here. For this reason, it makes sense to do the maximum amount of processing on the memristor. It seems that the amount of processing is limited by the length of the short-term memory of the memristor (which we suspect is related to the material properties, so can be tuned by materials scientists and chemists) and our instrumentation: the speed and accuracy of input and the accuracy of determining output. 

The use of memristor summation approaches in the full adder scheme is similar to how neurons work. For example three $\1$ inputs received one after the other causes the largest response spike and the only positive $t_2$ spike, either of these outputs could be linked to a threasholded switch which could release a current or voltage spike and thus act like a leaky integrate and fire neuron. The diminishing returns effect could enforce a refractory period. As neurons work by converting a rate-coded spiking voltage to a current spike at the synapse and then to a voltage spike, all of which can be considered transmission of a logical $\1$, the memristor with its action whereby input and output are current and voltage, could be ideally suited to neuromorphic computing.

Spiking logic is more brain-like. The brain is a very complex and not very well understood biological machine, but it is known that it operates slower than modern electronics and via spikes (neural spikes are voltage spikes caused by an influx in current rather than current spikes caused by a change in voltage, but the dynamics are very similar). The fact that memristors implement spiking logic naturally supports Chua's thesis that sodium and potassium ion gate proteins in the neural membranes are memristors and the habituation experimental data in section~\ref{sec:Bio} provides further evidence.


\subsection{Future Work}

This work suggests the following design for a memristor computer. A base level of memristors performing operations utilising the time-complexity as outlined here (i.e. interaction of several bits through time). The level above this could use space complexity whereby extra information is stored in how the memristor computing units are wired together. Add in some plasticity within and between the layers to allow learning and development of the system and we have a good model for a neuromorphic computer. This can be also be read as model of the brain, and it is my hope that neuroscientists investigate this point of view to see if it is useful to them. if the model fits, it would provide further evidence that Chua is correct in his ideas that memristors are related to neural circuits.


I strongly suspect that the Universial Memristor Machine~\cite{368}, if built with spiking memristor logic, would be a step change in neuromorphic computing, as this approach would combine the extra functionality due to the memory in processing and also circuit layout. Bottom layer: Memristor circuitry operating via memrory and ustilising time complexity. Top layer: space complexity, things wired together around it between the two: plasticity to allow for rewiring things on the fly. This is basically a description of the brain. And perhaps, adding in the trinary ideal gates would make it even more efficient and powerful.

\section*{Acknowledgment}
The author is indebted to Leon Chua for pointing out that her experimental results were the analogue of his theoretical modelling of Kandel's \emph{Aplysia} experiments and his enthusiasn for this work that took this paper out of convalescence. She is also indebted to Andrew Adamatzky for explaining the existence of sequential, asynchronous and trinary logics, Oliver Matthews and Alaric Snell-Pym  for computer science discussions, Ben de Lacy Costello for critical discussions and Leon Chua, Andrew Adamatkzky, Deborah Gater and Oliver Matthews for critical pre-review of this paper.

\bibliographystyle{IEEEtran.bst}

\maketitle
\bibliography{./UWELit}

\begin{thebibliography}{10}
\providecommand{\url}[1]{#1}
\csname url@samestyle\endcsname
\providecommand{\newblock}{\relax}
\providecommand{\bibinfo}[2]{#2}
\providecommand{\BIBentrySTDinterwordspacing}{\spaceskip=0pt\relax}
\providecommand{\BIBentryALTinterwordstretchfactor}{4}
\providecommand{\BIBentryALTinterwordspacing}{\spaceskip=\fontdimen2\font plus
\BIBentryALTinterwordstretchfactor\fontdimen3\font minus
  \fontdimen4\font\relax}
\providecommand{\BIBforeignlanguage}[2]{{%
\expandafter\ifx\csname l@#1\endcsname\relax
\typeout{** WARNING: IEEEtran.bst: No hyphenation pattern has been}%
\typeout{** loaded for the language `#1'. Using the pattern for}%
\typeout{** the default language instead.}%
\else
\language=\csname l@#1\endcsname
\fi
#2}}
\providecommand{\BIBdecl}{\relax}
\BIBdecl

\bibitem{14}
L.~O. Chua, ``Memristor - the missing circuit element,'' \emph{IEEE Trans.
  Circuit Theory}, vol.~18, pp. 507--519, 1971.

\bibitem{72}
M.~D. Ventra, Y.~V. Pershin, and L.~O. Chua, ``Putting memory into circuit
  elements: memristors, memcapcitors and meminductors.''

\bibitem{84}
L.~O. Chua and S.~M. Kang, ``Memristive devices and systems,''
  \emph{Proceedings of the IEEE}, vol.~64, pp. 209--223, 1976.

\bibitem{119}
L.~Chua, ``Resistance switching memories are memristors,'' \emph{Applied
  Physics A: Materials Science \& Processing}, pp. 765--782, 2011.

\bibitem{RevMemReRAM}
E.~Gale, ``Memristors and reram: Materials, mechanisms and models (a review),''
  \emph{Semiconductor Science and Technology}, vol.~29, p. 104004, 2014.

\bibitem{222}
H.~Kim, M.~P. Sah, and S.~P. Adhikari, ``Pinched hysteresis loops is the
  fingerprint of memristive devices.''

\bibitem{15}
D.~B. Strukov, G.~S. Snider, D.~R. Stewart, and R.~S. Williams, ``The missing
  memristor found,'' \emph{Nature}, vol. 453, pp. 80--83, 2008.

\bibitem{M0}
E.~Gale, R.~Mayne, A.~Adamatzky, and B.~de~Lacy~Costello, ``Drop-coated
  titanium dioxide memristors,'' \emph{Materials Chemistry and Physics}, vol.
  143, pp. 524--529, January 2014.

\bibitem{260}
\BIBentryALTinterwordspacing
E.~Gale, D.~Pearson, S.~Kitson, A.~Adamatzky, and B.~de~Lacy~Costello, ``The
  effect of changing electrode metal on solution-processed flexible titanium
  dioxide memristors,'' \emph{Materials Chemistry and Physics}, vol. 162, pp.
  20 -- 30, 2015. [Online]. Available:
  \url{http://www.sciencedirect.com/science/article/pii/S0254058415002072}
\BIBentrySTDinterwordspacing

\bibitem{296}
A.~Sawa, ``Resistive switching in transition metal oxides,'' \emph{Materials
  Today}, vol.~11, pp. 28--36, 2008.

\bibitem{155}
R.~Waser and M.~Aono, ``Nanoionics-based resistive switching memories,''
  \emph{Nature Materials}, vol.~6, pp. 833--840, 2007.

\bibitem{hystc}
E.~Gale, B.~de~Lacy~Costello, V.~Erokhin, and A.~Adamatzky, ``The short-term
  memory (d.c. response) of the memristor demonstrates the causes of the
  memristor frequency effect,'' in \emph{Proceedings of CASFEST 2014}, June
  2014.

\bibitem{SpcJ}
E.~Gale, B.~de~Lacy~Costello, and A.~Adamatzky, ``Observation, characterization
  and modeling of memristor current spikes,'' \emph{Appl. Math. Inf. Sci.},
  vol.~7, pp. 1395--1403, 4,July 2013.

\bibitem{123}
J.~Wu, K.~Mobley, and R.~L. McCreery, ``Electronic characteristics of fluorene
  / tio$_2$ molecular heterojunctions,'' \emph{The Journal of Chemical
  Physics}, vol. 126, p. 024704, 2007.

\bibitem{macro1}
W.~Alan~Doolittle, W.~Calley, and W.~Henderson, ``Complementary oxide memristor
  technology facilitating both inhibitory and excitatory synapses for potential
  neuromorphic computing applications,'' in \emph{Semiconductor Device Research
  Symposium, 2009. ISDRS '09. International}, Dec 2009, pp. 1--2.

\bibitem{261}
S.~Williams, ``Dynamics and applications of non-volatile and locally active
  memristors,'' June 2012, talk.

\bibitem{MooresLaw}
G.~E. Moore, ``Cramming more components onto integrated circuits,''
  \emph{Electronics Magazine}, p.~4, 1965.

\bibitem{322}
\BIBentryALTinterwordspacing
``International technology roadmap for semiconductors 2012 update overview,''
  International Technology Roadmap for Semiconductors 2012, Tech. Rep., 2012.
  [Online]. Available: \url{http://www.itrs.net/Links/2012ITRS/Home2012.htm}
\BIBentrySTDinterwordspacing

\bibitem{247}
L.~Chua, V.~Sbitnev, and H.~Kim, ``Hodgkin-{Huxley} axon is made of
  memristors,'' \emph{International Journal of Bifurcation and Chaos}, vol.~22,
  p. 1230011 (48pp), 2012.

\bibitem{248}
------, ``Neurons are poised near the edge of chaos,'' \emph{International
  Journal of Bifurcation and Chaos}, vol.~11, p. 1250098 (49pp), 2012.

\bibitem{41}
B.~Linares-Barranco and T.~Serrano-Gotarredona, ``Memristance can explain
  spike-time dependent plasticity in neural synapses,'' \emph{Nature
  Precedings}, 2009.

\bibitem{217}
V.~Erokhin, T.~Berzina, P.~Camorani, A.~Smerieri, D.~Vavoulis, J.~Feng, and
  M.~P. Fontana, ``Material memristive device circuits with synaptic
  plasticity: Learning and memory,'' \emph{BioNanoSci}, pp.
  DOI:10.1007/s12\,668--011--0004--7, April 2011.

\bibitem{272}
G.~M.-R. Matthew D.~Pickett and R.~S. Williams, ``A scalable neuristor built
  with mott memristors,'' \emph{Nature Materials}, vol.~16, pp. 114--117, Dec
  2012.

\bibitem{256}
T.~Masumoto, ``A chaotic attractor from chua's circuit,'' \emph{IEEE Trans.
  Circuits and Systems}, vol. CAS-31, pp. 1055--1058, 1984.

\bibitem{257}
R.~N. Madan, \emph{Chua's Circuit: A Paradigm for Chaos}.\hskip 1em plus 0.5em
  minus 0.4em\relax World Scientific, Singapore, 1993.

\bibitem{252}
A.~Buscarino, L.~Fortuna, M.~Frasca, and L.~V. Gambuzza, ``A chaotic circuit
  based on {Hewlett-Packard} memristor,'' \emph{Chaos}, vol.~22, p. 023136,
  2012.

\bibitem{82}
M.~Messias, C.~Nespoli, and V.~A. Botta, ``Hopf bifurcation from lines of
  equilibria without parameters in memristor oscillators,'' \emph{International
  Journal of Bifurcation and Chaos}, vol.~20, pp. 437--450, 2010.

\bibitem{70}
B.-C. Bao, J.-P. Xu, and Z.~Liu, ``Initial state dependent dynamical behaviors
  in a memristor based chaotic circuit,'' \emph{Chinese Physics Letters},
  vol.~27, p. 070504, 2010.

\bibitem{61}
B.~C. Bao, Z.~Liu, and J.~P. Xu, ``Steady periodic memristor oscillator with
  transient chaotic behaviours,'' \emph{Electronics Letters}, pp. 237--238,
  2010.

\bibitem{232}
B.~Bo-Cheng, X.~J. Ping, Z.~Guo-Hua, M.~Zheng-Hua, and Z.~Ling, ``Chaotic
  memristive circuit: equivalent circuit realization and dynamical analysis,''
  \emph{Chinese Physics B}, vol.~20, p. 120502 (7pp), 2011.

\bibitem{EllaMattia}
M.~F. Lucia Valentina~Gabuzza, Luigi~Fortuna and E.~Gale, ``Experimental
  evidence of chaos from memristors,'' \emph{International Journal of
  Bifurcation and Chaos}, vol. (forthcoming), 2015.

\bibitem{EllaC1}
E.~Gale, B.~de~Lacy~Costello, and A.~Adamatzky, ``Emergent spiking in non-ideal
  memristor networks,'' \emph{Microelectronics Journal}, vol.~45, pp.
  1401--1415, 2014.

\bibitem{HH}
A.~L. Hodgkin and A.~F. Huxley, ``A quantitative description of membrane
  current and its application to conduction in nerve,'' \emph{J. Physiol.},
  vol. 117, pp. 500--544, 1952.

\bibitem{ChuaNanotech}
\BIBentryALTinterwordspacing
L.~Chua, ``Memristor, hodgkin–huxley, and edge of chaos,''
  \emph{Nanotechnology}, vol.~24, no.~38, p. 383001, 2013. [Online]. Available:
  \url{http://stacks.iop.org/0957-4484/24/i=38/a=383001}
\BIBentrySTDinterwordspacing

\bibitem{SquireMemoryBook}
\BIBentryALTinterwordspacing
L.~Squire and E.~Kandel, \emph{Memory: From Mind to Molecules}, ser. Owl
  book.\hskip 1em plus 0.5em minus 0.4em\relax Henry Holt and Company, 2000.
  [Online]. Available: \url{http://books.google.ae/books?id=pqoYubI2GhsC}
\BIBentrySTDinterwordspacing

\bibitem{374}
E.~Antonova, P.~Chadwick, and V.~Kumari, ``'more meditation, less habituation?
  the effect of mindfulness practice on the acoustic startle reflex.',''
  \emph{PLoS ONE}, vol.~10, p. 0123512, 2015.

\bibitem{adamatzky2007unconventional}
A.~Adamatzky, L.~Bull, and B.~D.~L. Costello, \emph{Unconventional computing
  2007}.\hskip 1em plus 0.5em minus 0.4em\relax Luniver Press, 2007.

\bibitem{adamatzky2006utopian}
A.~Adamatzky and C.~Teuscher, \emph{From utopian to genuine unconventional
  computers}.\hskip 1em plus 0.5em minus 0.4em\relax Luniver Press, 2006.

\bibitem{gheorghe2005molecular}
M.~Gheorghe, \emph{Molecular Computational Models: Unconventional
  Approaches}.\hskip 1em plus 0.5em minus 0.4em\relax IGI Global, 2005.

\bibitem{AndysBook}
A.~Adamatzky, \emph{Physarum Machines: Computers from Slime Mould}, ser. World
  Scientific Series on Nonlinear Science Series A.\hskip 1em plus 0.5em minus
  0.4em\relax Prentice-Hall, Upper Saddle River, NJ, 1994, vol.~74.

\bibitem{adamatzky2001computing}
------, \emph{Computing in nonlinear media and automata collectives}.\hskip 1em
  plus 0.5em minus 0.4em\relax CRC Press, 2001.

\bibitem{377}
M.~Karnaugh, ``The map method for synthesis of combinational logic circuits,''
  \emph{Transactions of the American Institute of Electrical Engineers part I},
  vol.~72, pp. 593--599, 1953.

\bibitem{378}
E.~W.~. Veitch., ``Chart method for simplifying truth functions,''
  \emph{Transactions of the 1952 ACM Annual Meeting, ACM Annual
  Conference/Annual Meeting}, pp. 127--133, 1952.

\bibitem{Shannon}
C.~Shannon, ``A symbolic analysis of relay and switching circuits,''
  \emph{Trans. AIEE}, vol.~57, pp. 713--723, 1938.

\bibitem{242}
J.~Borghetti, G.~D. Snider, P.~J. Kuekes, J.~J. Yang, D.~R. Stewart, and R.~S.
  Williams, ```memristive' switches enable `stateful' logic operations via
  material implication,'' \emph{Nature}, vol. 464, pp. 873--876, 2010.

\bibitem{370}
E.~Lehtonen and M.~Laiho, ``Stateful implication logic with memristors,'' in
  \emph{Nanoscale Architectures, 2009. NANOARCH '09. IEEE/ACM International
  Symposium on}, July 2009, pp. 33--36.

\bibitem{371}
E.~Lehtonen, J.~Poikonen, and M.~Laiho, ``Applications and limitations of
  memristive implication logic,'' in \emph{Cellular Nanoscale Networks and
  Their Applications (CNNA), 2012 13th International Workshop on}, Aug 2012,
  pp. 1--6.

\bibitem{243}
E.~M. Gale, B.~de~Lacy~Costello, and A.~Adamatzky, ``Observation and
  characterization of memristor current spikes and their application to
  neuromorphic computation,'' in \emph{2012 International Conference on
  Numerical Analysis and Applied Mathematics (ICNAAM 2012)}, Kos, Greece, Sept
  2012.

\bibitem{Mem1}
R.~Thompson and W.~Spencer, ``Habituation: A model phenomena for the study of
  neural substrates of behaviour,'' \emph{Psychological Review}, vol. 173, pp.
  16--43, 1966.

\bibitem{Mem2}
C.~Bailey and M.~Chen, ``Morphological basis of long-term habituation and
  sensitisation in \emph{Aplysia},'' \emph{Science}, vol. 220, pp. 91--93,
  1983.

\bibitem{6815415}
V.~Erokhin, ``Organic memristive devices: Architecture, properties and
  applications in neuromorphic networks,'' in \emph{Electronics, Circuits, and
  Systems (ICECS), 2013 IEEE 20th International Conference on}, Dec 2013, pp.
  305--308.

\bibitem{C2JM35064E}
\BIBentryALTinterwordspacing
V.~Erokhin, T.~Berzina, K.~Gorshkov, P.~Camorani, A.~Pucci, L.~Ricci,
  G.~Ruggeri, R.~Sigala, and A.~Schuz, ``Stochastic hybrid 3d matrix: learning
  and adaptation of electrical properties,'' \emph{J. Mater. Chem.}, vol.~22,
  pp. 22\,881--22\,887, 2012. [Online]. Available:
  \url{http://dx.doi.org/10.1039/C2JM35064E}
\BIBentrySTDinterwordspacing

\bibitem{P0c}
E.~Gale, B.~de~Lacy~Costello, and A.~Adamatzky, ``Boolean logic gates from a
  single memristor via low-level sequential logic,'' in \emph{Submitted}.

\bibitem{376}
R.~E. Simpson, Reidel, Ed.\hskip 1em plus 0.5em minus 0.4em\relax Allyn and
  Bacon, Boston, 1987.

\end{thebibliography}
%



\end{document}